\documentstyle[12pt,epsf]{article}
\newlength{\mytopmargin}
\newlength{\myleftmargin}
\renewcommand{\theequation}{\thesection.\arabic{equation}}

\newcommand{\beqa}{\begin{eqnarray}}
\newcommand{\eeqa}{\end{eqnarray}}

\begin{document}
\title{\Large\bf Correlations for the orthogonal-unitary \\
and symplectic-unitary transitions at \\
the hard and soft edges}
\author{P.J.~Forrester$^{\dagger}$, T.~Nagao$^{*}$ and 
G.~Honner$^{\dagger}$}

\date{}
\maketitle
\begin{center}
 $^{\dagger}$Department of Mathematics, University of Melbourne, \\
 Parkville, Victoria 3052, Australia
\end{center}
\begin{center}
$^*$
Department of Physics, Faculty of Science, Osaka University, \\
Toyonaka, Osaka 560, Japan
\end{center}

\bigskip
\begin{center}
\bf Abstract
\end{center}
\par
\bigskip
\noindent

For the orthogonal-unitary and symplectic-unitary transitions in random
matrix theory, the general parameter dependent distribution between
two sets of eigenvalues with two different parameter values can be
expressed as a quaternion determinant. For the parameter dependent
Gaussian and Laguerre ensembles the matrix elements of the determinant
are expressed in terms of corresponding skew-orthogonal polynomials,
and their limiting value for infinite matrix dimension are
computed in the vicinity of the soft and hard edges respectively.
A connection formula relating the distributions at the hard and soft
edge is obtained, and a universal asymptotic behaviour of the two point
correlation is identified.
\section{Introduction}
Random matrix ensembles for quantum Hamiltonians are defined in terms
of the constraint imposed on the general Hermitian matrix representing
the Hamiltonian by time reversal symmetry and spin-rotation invariance
(when relevant). This leads to ten distinct random matrix ensembles
\cite{Zi97}.  The corresponding eigenvalue 
probability 
density function (p.d.f.) can be computed exactly when the independent 
elements are 
chosen to have a Gaussian distribution. One then finds that the ten 
distinct 
random matrix ensembles belong to one of six different eigenvalue p.d.f.'s. 
These are 
\beqa 
\label{1.1}
{1 \over C_N} \prod_{l=1}^N e^{-\beta x_l^2/2} \prod_{1 \le j < k \le N}
|x_k - x_j|^\beta
\eeqa 
which defines the Gaussian random matrix ensemble, or 
\beqa
\label{1.2}
{1 \over C_N'} \prod_{l=1}^N e^{-\beta x_l^2/2} 
x_l^{\beta a + 1} \prod_{1 \le j < k \le N}
|x_k^2 - x_j^2|^\beta, \quad x_l >0
\eeqa
which defines the Laguerre random matrix ensemble, where $\beta = 1, 2$ or 4. 

The study of these random matrix ensembles can be generalized to include a 
parameter $\tau$. This allows the situation to be studied in which the system 
evolves between two distinct eigenvalue p.d.f.'s as $\tau$ is varied. For the 
Gaussian ensemble this can be achieved by choosing the joint 
distribution of the 
elements of the random matrix to be
\beqa
\label{1.3}
P({X}^{(0)};{X};\tau) = A_{\beta,\tau}
\exp\Big (-\beta
{\rm Tr} \Big \{ (
{X} - {\rm e}^{-\tau}{X}^{(0)})^2
\Big \}/2|1 - {\rm e}^{-2 \tau}| \Big ).
\eeqa
In this expression $X$ is a real symmetric 
($\beta = 1$), Hermitian ($\beta = 2$) 
or self dual quaternion ($\beta = 4$) $N \times N$ matrix, and $X^{(0)}$ is a 
prescribed random matrix which must belong to a subspace of the symmetry 
class of $X$. For $\tau \rightarrow \infty$ the p.d.f. is independent of 
$X^{(0)}$ and is that of the standard Gaussian ensembles, with p.d.f. 
(\ref{1.1}). For general $\tau$ 
Dyson \cite{Dy63} proved that the eigenvalue p.d.f. 
$p = p(x_{1},  \ldots , x_{N}; \tau)$ satisfies the Fokker-Planck equation 
\beqa
\label{1.4}
{\partial p \over \partial \tau} = {\cal L}p, \quad
{\cal L} = {1 \over \beta} \sum_{j=1}^N {\partial \over \partial x_j}
e^{-\beta W} {\partial \over \partial  x_j}  e^{\beta W}
\eeqa
with 
\beqa
\label{1.5}
W = W^{(H)}   =  {1 \over 2} \sum_{j=1}^N x_j^2 - \sum_{1 \le j < k \le N}
\log |x_k - x_j|
\eeqa
(the superscript $(H)$ is used because of a connection with the Hermite 
polynomials), 
subject to the initial condition that $p$ agrees with the eigenvalue 
p.d.f.~of $X^{(0)}$ at $\tau = 0$. 

For the Laguerre ensemble, a parameter dependent theory can be developed 
(see e.g.~\cite{Fo98}) by 
considering the non-negative matrices $A = X^{\dagger} X$, where $X$ is an 
$n \times m$ parameter-dependent matrix 
\beqa
\label{1.6}
P({X}^{(0)};{X};\tau) = A_{\beta,\tau}
\exp\Big (-\beta
{\rm Tr} \Big \{ (
{X} -  e^{-\tau}{X}^{(0)})^\dagger   (
{X} -  e^{-\tau}{X}^{(0)})
\Big \}/2|1 - {\rm e}^{-2 \tau}| \Big )
\eeqa
(c.f.\ (\ref{1.3})) and the 
elements of $X$ are real ($\beta = 1$), complex ($\beta = 2$)  
or quaternion 
real ($\beta = 4$), and  $X^{(0)}$ is a prescribed random matrix which 
must belong to a 
subspace of the symmetry class of $X$. For $\tau \rightarrow \infty$ 
the corresponding 
p.d.f.~for the square of the eigenvalues of $A$ is given by (\ref{1.2}) 
with $a = n - m + 1 - 2/\beta$, while for general $\tau$ the eigenvalue p.d.f. 
$p = p(x_{1}^{2},  \ldots , x_{N}^{2}; \tau)$ satisfies (\ref{1.4}) with 
\beqa
\label{1.7}
W = W^{(L)}  =  {1 \over 2} \sum_{j=1}^N x_j^2  - (a + 1/\beta) 
 \sum_{j=1}^N \log x_j - \sum_{1 \le j < k \le N}
 \log |x_k^2 - x_j^2|
\eeqa
(the superscript 
$(L)$ is used because of a connection with the Laguerre polynomials).

Associated 
with the parameter dependent eigenvalue p.d.f.'s are parameter dependent 
distribution functions 
\beqa
\label{1.7'}
\rho_{(n+m)}(x_1^{(1)},\dots,x_n^{(1)};\tau_1;x^{(2)}_1,\dots,x_m^{(2)};
\tau_2)
\eeqa
The distribution function $\rho_{(n+m)}$ is defined such that the ratio 
\beqa
\nonumber
\rho_{(n+m)}(x_1^{(1)},\dots,x_n^{(1)};\tau_1;x_1^{(2)},\dots,x_m^{(2)};
\tau_2)/\rho_{(n+(m-1))}(x_1^{(1)},
\dots,x_n^{(1)};\tau_1;x^{(2)},\dots,x_{m-1}^{(2)};
\tau_2)
\eeqa
gives the density 
of eigenvalues at $x_{m}^{(2)}$ when the parameter  equals $\tau_{2}$, 
given that there 
were eigenvalues at $x_{1}^{(1)}, \ldots,  x_{n}^{(1)}$ when the 
parameter equalled $\tau_{1}$, and that there are also eigenvalues at 
$x_{1}^{(2)}, \ldots,  x_{m}^{(2)}$ with parameter $\tau_{2}$ (it is assumed 
$\tau_{2} > \tau_{1}$). 

For the parameter 
dependent Gaussian and Laguerre ensembles, quaternion determinant expressions 
have been 
derived for $\rho_{(n+m)}$ in the case that $\beta = 2$ in (\ref{1.4}) and 
initial condition 
proportional to $e^{-\beta ' W}$ with $\beta ' = 1$ and 4 \cite{NF98a}. 
Because each 
$\beta$ corresponds to a different symmetry class of the Hamiltonian 
(orthogonal, 
unitary and symplectic symmetry for $\beta = 1, 2$ and 4 respectively), 
this describes the transition from orthogonal or symplectic symmetry to unitary 
symmetry as the parameter $\tau$ is varied from $0$ to $\infty$. 

The determinant 
expressions apply for the finite system. As revised in Section 2, they 
involve 
skew orthogonal polynomials known from the study of the distribution functions 
for $\tau = 0$ in the $\beta = 1$ and 4 cases \cite{NW91,NF95}, the orthogonal 
polynomials 
related to the $\tau = 0$, $\beta = 2$ case, as well as the elements of a 
connection matrix which specifies the orthogonal polynomials in terms of the 
appropriate 
skew orthogonal polynomials. In the Gaussian case, this matrix can be 
written 
down from knowledge of the expression for the skew orthogonal polynomials in 
terms of the 
orthogonal polynomials (the Hermite polynomials). In the Laguerre case this 
step requires some calculation which is undertaken below. 

Once all 
these quantities are specified, the general $(n + m)$-point distribution 
function in the finite system is known explicitly. 
The immediate task is to compute the scaled thermodynamic limit. This is
done in Section 3 for the Laguerre ensemble 
in the neighbourhood
of the hard edge (the smallest
eigenvalues, which are constrained to be positive), and in Section 4
for the Gaussian ensemble in the neighbourhood of the soft edge (the
largest eigenvalues).

In Section 5 inter-relationships between the distribution functions
are discussed. This is done by analyzing certain limits in (\ref{1.7'}):
the limit $\tau_1,\tau_2 \to \infty$, $\tau_1 - \tau_2$ fixed; the 
$a \to \infty$ limit of the hard edge; and the bulk limit corresponding to
large
distances from the edge. Also considered is the asymptotic form of
the dynamical density-density function $\rho_{(1+1)}^T(X,Y;t)$, which is
shown to exhibit a universal non-oscillatory decay for $X,Y,t \to \infty$.
A consequence of this decay is an O(1) fluctuation formula for the
variance of a slowly varying linear statistic. In the final subsections
a realization of the density at the soft edge for $\beta = 1$ initial
conditions is given by  an
empirical computation based on 
the parameter dependent matrices (\ref{1.3})
with $X^{(0)}$ a real symmetric matrix, and the explicit form of the
static distribution functions at the hard and soft edges for $\beta = 1$
and 4 is noted.

In the Appendix it is
shown how our results for the Laguerre ensemble with $N$ finite
coincide, in the static limit $\tau \to 0$, with formulas given recently
by Widom \cite{Wi98}.

\section{Revision of the formalism}
\setcounter{equation}{0}

At the 
special coupling $\beta = 2$, the Fokker-Planck equation (\ref{1.4}) with $W$ 
given by (\ref{1.5}) or (\ref{1.7}) can be transformed into an imaginary-time 
Schr\"{o}dinger 
equation for $N$ independent fermions (see e.g.\ \cite{Fo98}). As such 
the exact $N$-body 
Green function $G(\{x_{j}^{(0)}\}, \{x_{j}\}; \tau)$ can be written 
as a determinant 
involving the corresponding single particle Green functions. In terms of 
the new variable 
$y_{j} = x_{j}$ in (\ref{1.1}) and $y_{j} = x_{j}^{2}$ in (\ref{1.2}) we 
have previously shown \cite{NF98} that 
\beqa
\label{2.1}
G(x_1^{(0)},\cdots,x_N^{(0)};x_1,\cdots,x_N;\tau)  dx_1 \cdots
 dx_N
= G(y_1^{(0)},\cdots,y_N^{(0)};y_1,\cdots,y_N;\tau)  dy_1\cdots 
dy_N
\eeqa
where
\begin{eqnarray}
\label{2.2}
& & G(y_1^{(0)},\cdots,y_N^{(0)};y_1,\cdots ,y_N;\tau) \nonumber \\ 
& = & {\rm e}^{E_0\tau/2} 
\prod_{j=1}^N \sqrt{w(y_j) \over w(y_j^{(0)})} \prod_{1 \le j < k \le N} 
{(y_k - y_j) \over (y_k^{(0)}-y_j^{(0)})} \det[
g(y_j^{(0)},y_k;\tau)]_{j,k=1,
\cdots,N}
\end{eqnarray}
Here 
\beqa
\label{2.3}
g(y^{(0)},y;\tau) = \sqrt{w(y^{(0)})w(y)} \sum_{j=0}^\infty
p_j(y^{(0)})p_j(y) {\rm e}^{-\gamma_j\tau}
\eeqa
and $\{p_{j}(x)\}$ are the orthonormal polynomials 
associated with the weight function 
$w(x)$. In the Gaussian case 
\beqa
\label{2.4}
w(y)={\rm e}^{-y^2}, p_n(y) = 2^{-n/2} \pi^{-1/4} (n!)^{-1/2} H_n(y), \quad
\gamma_n = n + {1
\over 2},
\eeqa
where $H_n(y)$ denotes the Hermite polynomial,
while for the Laguerre ensemble 
\beqa
\label{2.5}
w(y) = y^{a} {\rm e}^{-y}, \quad
p_n(y) = (-1)^n \Big ( {\Gamma(n+1) \over
\Gamma(a+n+1)} \Big )^{1/2} L_n^a(y), \quad \gamma_n = 2(n + {a + 1 \over
2})
\eeqa
where $L_n^a(y)$ denotes the Laguerre polynomial.
The factor 
$e^{E_{0}\tau/2}$ cancels out the formula for the distribution function,  
so the explicit value of $E_{0}$ is not needed. 

Knowledge of the Green function allows the p.d.f. 
\beqa
\nonumber 
 p(y_1^{(1)},\cdots,y_N^{(2)};\tau_1;y_1^{(2)},\cdots ,y_N^{(2)};\tau_2)
\eeqa
for the event that there are eigenvalues at $y_{1}^{(1)}, \ldots , y_{N}^{(1)}$ when the 
parameter equals $\tau_{1}$, 
and at $y_{1}^{(2)}, \ldots , y_{N}^{(2)}$ when the 
parameter is increased to $\tau_{2}$ to be calculated. Explicitly 
\begin{eqnarray}
\label{2.a}
\lefteqn{ p(y_1^{(1)},\cdots,y_N^{(2)};\tau_1;y_1^{(2)},\cdots
,y_N^{(2)};\tau_2)} \nonumber
\\ && 
= {1 \over N!}
\int_{I'} dy_1^{(0)} \cdots \int_{I'} dy_N^{(0)}
\,p_0(y_1^{(0)},\cdots
,y_N^{(0)})
G(y_1^{(0)},\cdots,y_N^{(0)};y_1^{(1)},\cdots,y_N^{(1)};\tau_1) \nonumber
\\ \quad && \times
G(y_1^{(1)},\cdots,y_N^{(1)};y_1^{(2)},\cdots,y_N^{(2)};\tau_2 - \tau_1).
\end{eqnarray}
Here $p_{0}$ denotes the prescribed initial p.d.f. (in the $y$ variables) 
and $I'$ denotes 
the transformed domain
(for convenience this will be omitted from subsequent formulas).  
Initial p.d.f.'s with orthogonal ($\beta = 1$) and symplectic 
($\beta = 4$) symmetry are given by the functional forms 
\beqa 
\label{2.b}
p_0(y_1,\cdots,y_N) \propto \left\{
\begin{array}{ll} \prod_{j=1}^N \sqrt{w(y_j)} \prod_{j < l}^N |y_j - y_l|,
& \beta = 1 \\[.2cm]
 \prod_{j=1}^{N/2} w(y_j) \delta(y_j - y_{j+N/2}) \prod_{j < l}^{N/2}
 |y_j - y_l|^4,
& \beta = 4 \quad  (N\: {\rm even}), 
\end{array}
\right.
\eeqa

The parameter dependent distribution function $\rho_{(n+m)}$ is calculated from the 
p.d.f.~(\ref{2.a}) according to the formula 
\begin{eqnarray}
\label{2.c}
\lefteqn{
\rho_{(n+m)}(y_1^{(1)},\cdots,y_N^{(1)};\tau_1;
y_1^{(2)},\cdots,y_N^{(2)};\tau_2) =
   \frac{N!}{(N-n)!}{N! \over (N - m)!}} \nonumber \\&&
\times \int dy_{n+1}^{(1)} \cdots \int dy_N^{(1)}
\int dy_{m+1}^{(2)} \cdots \int dy_N^{(2)} 
 p(y_1^{(1)},\cdots,y_N^{(2)};\tau_1;y_1^{(2)},\cdots ,y_N^{(2)};\tau_2)
\end{eqnarray}
In a recent 
work \cite{NF98a} we have shown that for initial conditions (\ref{2.b}), 
and $G$ specified by (\ref{2.2}) 
with general weight function $w(y)$ and corresponding 
orthogonal polynomials 
$\{p_{k}(y)\}_{k = 0, 1, \ldots}\,$, the quantity (\ref{2.c}) 
can be expressed as a $(n+m)$-dimensional quaternion determinant Tdet (this 
quantity, which was introduced into random matrix theory by Dyson \cite{Dy70}, 
is explicitly defined in e.g.~\cite{NF98}). 

The entries of the quaternion determinant depend on the functions 
\beqa 
\nonumber
S(x,y;\tau_x,\tau_y), \quad \tilde{I}(x,y;\tau_x,\tau_y), \quad
D(x,y;\tau_x,\tau_y).
\eeqa
These functions in turn depend on 
the quantity $F(x, y; \tau)$, which is defined for 
the $\beta = 1$ initial conditions in (\ref{2.b}) by 
\beqa
\label{2.d}
F(x,y;\tau)  =  \int  dz^{\prime} \int^{z^{\prime}} dz
\Big ( g(x,z;\tau) g(y,z^{\prime};\tau) -
g(y,z;\tau) g(x,z^{\prime};\tau) \Big ), \label{tu.1}
\eeqa
while for the $\beta = 4$ initial conditions in (\ref{2.b}) it is specified by 
\beqa 
\label{2.e}
F(x,y;\tau)  =  \int  dz  \{ g(x,z;\tau)
\frac{\partial}{\partial z} g(y,z;\tau) - g(y,z;\tau)
\frac{\partial}{\partial z}  g(x,z;\tau) \}. \label{tu.2}
\eeqa

They also depend 
on a family of skew-orthogonal monic polynomials $R_{k}(x; \tau)$ 
of degree $k$. The terminology skew-orthogonal means that they satisfy the 
skew orthogonality relations 
\begin{eqnarray} 
\label{2.f}
 < R_{2m}(\cdot;\tau), R_{2n+1}(\cdot;\tau) > =
 - < R_{2n+1}(\cdot;\tau), R_{2m}(\cdot;\tau)> = r_m(\tau) \delta_{m,n},
\nonumber
\\
 < R_{2m}(\cdot;\tau), R_{2n}(\cdot;\tau) > = 0, \quad
  < R_{2m+1}(\cdot;\tau), R_{2n+1}(\cdot;\tau) > = 0
\end{eqnarray}
where 
\beqa 
\label{2.g}
 < f(x), g(y) > = \frac{1}{2} \int  dx \int  dy
 \sqrt{w(x) w(y)} F(y,x;\tau)  \Big ( f(y) g(x) - f(x) g(y)\Big ). 
\eeqa
The polynomials $R_{k}(x; \tau)$ can be deduced from their static
counterparts   
\cite{FP95,NF98}. 
Suppose at $\tau = 0$ we write 
\beqa 
\label{2.h}
R_n(x;0) = \sum_{j=0}^n \alpha_{nj}C_j(x)
, \quad \alpha_{nn}=1,
\eeqa
where the $C_{j}(x)$ are the monic version of the orthogonal 
polynomials  $p_{j}(x)$ 
in (\ref{2.3}). Then it is straightforward to verify that 
\beqa 
\label{2.i}
R_n(x;\tau)  e^{\gamma_n \tau} = \sum_{j=0}^n \alpha_{nj} C_j(x)
e^{\gamma_j \tau}
\eeqa
and 
\beqa 
\label{2.j}
r_k(\tau) = r_k(0) e^{-(\gamma_{2k} + \gamma_{2k+1})\tau}
\eeqa

Still another function involved is 
\beqa 
\Phi_k(x;\tau) := \int F(y,x;\tau) \sqrt{w(y)} R_k(y;\tau) \,
 dy.
\label{2.k}
\eeqa
Now, if we invert (\ref{2.h}) by writing 
\beqa 
\label{2.l}
C_n(x)  e^{\gamma_n \tau} = \sum_{j=0}^n \beta_{nj}
e^{\gamma_j \tau} R_j(x;\tau), \ \ \beta_{nn} = 1.
\eeqa
then $\Phi_{k}(x; \tau)$ can be expanded according to \cite{NF98} 
\begin{eqnarray}\label{2.m}
\Phi_{2k-1}(x;\tau) {\rm e}^{\gamma_{2k-1} \tau} & = &
- \sqrt{w(x)}  r_{k-1}(0) \sum_{\nu = 2 k - 2}^{\infty} \frac{C_{\nu}(x)
e^{- \gamma_{\nu} \tau}}{h_{\nu}} \beta_{\nu \ 2 k - 2},
\nonumber
\\
\Phi_{2k-2}(x;\tau)  e^{\gamma_{2k-2} \tau}& = & \sqrt{w(x)}  r_{k-1}(0)
\sum_{\nu
= 2
k - 1}^{\infty} \frac{C_{\nu}(x)
e^{- \gamma_{\nu} \tau}}{h_{\nu}} \beta_{\nu \ 2 k - 1} .
\end{eqnarray}
where $h_{\nu}$ is the normalization 
\beqa 
\label{2.n}
\int w(x) C_j(x) C_k(x) \, dx = h_j \delta_{j,k}.
\eeqa

With this notation, we can now define the functions $S, \tilde{I}$ 
and $D$. We have, 
assuming $N$ even, 
\begin{eqnarray} 
\label{2.o} 
\lefteqn{
S(x,y;\tau_x,\tau_y)  :=   \sqrt{w(y)}} \nonumber \\
& & \times 
\sum_{l=0}^{N/2-1} \frac{e^{\gamma_{2l}\tau_x + \gamma_{2l+1}\tau_y}}
{r_{l}(0)}
\{ \Phi_{2 l }(x;\tau_x)  R_{2 l + 1}(y;\tau_y) -
\Phi_{2 l + 1}(x;\tau_x)  R_{2 l}(y;\tau_y) \}, 
\end{eqnarray}
\begin{eqnarray}
\label{2.p} \lefteqn{
\tilde{I}(x,y;\tau_x,\tau_y)} \nonumber \\
  &  & :=
\sum_{l=N/2}^\infty  \frac{e^{\gamma_{2l}\tau_x + \gamma_{2l+1}\tau_y}}
{r_{l}(0)}
\{ \Phi_{2 l}(x;\tau) \Phi_{2 l + 1}(y;\tau) -
\Phi_{2 l + 1}(x;\tau) \Phi_{2 l}(y;\tau) \}
\end{eqnarray}
\begin{eqnarray}
\label{2.q}\lefteqn{ 
D(x,y;\tau_x,\tau_y)  :=   \sqrt{w(x)w(y)} } \nonumber \\ && \times
\sum_{l=0}^{N/2-1} \frac{e^{\gamma_{2l}\tau_x + \gamma_{2l+1}\tau_y}}
{r_{l}(0)}
\{ R_{2 l }(x;\tau) R_{2 l + 1}(y;\tau) -
R_{2 l + 1}(x;\tau) R_{2 l }(y;\tau) \} 
\end{eqnarray}
and in terms of $S$ we define 
\beqa 
\label{2.r}
\tilde{S}(x,y;\tau_x,\tau_y) := S(x,y;\tau_x,\tau_y) - g(x,y;\tau_x-\tau_y),
\quad \tau_x > \tau_y.
\eeqa
The formula for $\rho_{(n+m)}$ can now be stated. It reads \cite{NF98a}
\begin{eqnarray} 
\lefteqn{\rho_{(n+m)}(x_1^{(1)},\cdots,x_N^{(1)};\tau_1;
x_1^{(2)},\cdots,x_N^{(2)};\tau_2)} \nonumber \\ && = {\rm Tdet}
\left [ \begin{array}{cc} [f_1(x_j^{(1)},x_k^{(1)};\tau_1,\tau_1)
]_{n \times n}
& [f_2(x_j^{(1)},x_k^{(2)};\tau_1,\tau_2)]_{n \times m} \\{}
 [f_2^D(x_j^{(1)},x_k^{(2)};\tau_1,\tau_2)]_{n \times m}^T & 
 [f_1(x_j^{(2)},x_k^{(2)};\tau_2,\tau_2)
 ]_{m \times m} \end{array} \right ]
\label{2.s}
\end{eqnarray}
where 
$$
f_1(x,y;\tau,\tau) := \left [ \begin{array}{cc}
S(x,y;\tau,\tau) & \tilde{I}(x,y;\tau,\tau) \\
D(x,y;\tau,\tau) & S(y,x;\tau,\tau) \end{array} \right ],
$$
$$ 
f_2(x,y;\tau_x,\tau_y) := \left [ \begin{array}{cc}
S(x,y;\tau_x,\tau_y) & \tilde{I}(x,y;\tau_x,\tau_y) \\
D(x,y;\tau_x,\tau_y) & \tilde{S}(y,x;\tau_y,\tau_x) \end{array} \right ]
$$
and
$$
f_2^D(x,y;\tau_x,\tau_y) := \left [ \begin{array}{cc}
\tilde{S}(y,x;\tau_y,\tau_x) & - \tilde{I}(x,y;\tau_x,\tau_y) \\
-D(x,y;\tau_x,\tau_y) & S(x,y;\tau_x,\tau_y) \end{array} \right ].
$$
%% Part 3; Page 10 starts here. 
If we denote the matrix in (\ref{2.s}) by $X$, and note from the definitions 
(\ref{2.p}) and (\ref{2.q}) that 
\beqa
\label{2.1a}
 \tilde{I}(x,y;\tau_x,\tau_y) = -   \tilde{I}(y,x;\tau_y,\tau_x), \quad
 D(x,y;\tau_x,\tau_y) = - D(y,x;\tau_y,\tau_x)
\eeqa
then we see that $ZX$ where 
\beqa
\nonumber
Z = {\bf 1}_{(n+m)} \otimes \left ( \begin{array}{cc}
0 & -1 \\ 1 & 0 \end{array} \right )
\eeqa
is an even dimensional antisymmetric matrix. Its Pfaffian is thus well defined, 
and in fact \cite{Dy70} 
\beqa
\label{2.1b}
{\rm Tdet} X = {\rm Pf}(ZX).
\eeqa
This formula can be used to compute the formula (\ref{2.s}). 

\subsection{Alternative formulas and inter-relationships}

As first noted in \cite{NF98}, 
substitution of (\ref{2.m}) in (\ref{2.o}) and use of 
(\ref{2.l}) implies the decomposition 
\beqa
\label{2.1c}
S(x,y;\tau_x,\tau_y) = S_1(x,y;\tau_x,\tau_y) +  S_2(x,y;\tau_x,\tau_y)
\eeqa
where 
\begin{eqnarray}
\label{2.1d} 
 S_1(x,y;\tau_x,\tau_y) & = &
 \sqrt{w(x) w(y)} \sum_{\nu = 0}^{N-1}
  \frac{C_{\nu}(x) C_{\nu}(y)e^{-\gamma_\nu(\tau_x-\tau_y)}}{h_{\nu}}
\\
\label{2.1e}
S_2(x,y;\tau_x,\tau_y) & = &  \sqrt{w(x) w(y)}
\sum_{\nu = N}^{\infty} \sum_{k=0}^{N-1}
\frac{C_{\nu}(x)  e^{-\gamma_{\nu} \tau_x}}{h_{\nu}} 
\beta_{\nu k}
R_k(y;\tau_y
)
{\rm e}^{\gamma_k \tau_y}
\end{eqnarray}

Since $0 \leq \gamma_{1} < \gamma_{2} < \cdots$ (recall (\ref{2.4}) and 
(\ref{2.5})) we see that for $N \rightarrow \infty$ and with all other 
parameters fixed, $S_{2} \rightarrow 0$ while 
\beqa
S_{1}\left(x, y; \tau_{x}, \tau_{y}\right) \rightarrow 
g\left(x, y; \tau_{x}-\tau_{y}\right) \, ,
\nonumber
\eeqa
assuming $\tau_{x} - \tau_{y} > 0$. The use of this result is that it shows 
\beqa
\label{2.1f}
\tilde{I}(x,y;\tau_x,\tau_y) = \int F(y',y;\tau_y) \Big (
g(x,y';\tau_x - \tau_y) - S(x,y';\tau_x,\tau_y) \Big ) \, dy'
\eeqa
valid for $\tau_{x} - \tau_{y} > 0$ (for $\tau_{x} - \tau_{y} < 0$ we can first use 
the first equation in (\ref{2.1a}), then apply (\ref{2.1f})). 

We would also like to relate $D$ to $S$.  Apparently this can be done naturally 
only when $\tau_{x} = 0$. 
Using the fact that $g(x, y; 0) = \delta(x - y)$ we see from 
(\ref{2.d}) and (\ref{2.e}) that 
\beqa
\label{2.1g}
F(x,y;0) = \left \{ \begin{array}{ll} {\rm sgn}(y-x), & \beta = 1 \\
2 {\partial \over \partial x} \delta (y-x), & \beta = 4. \end{array}
\right.
\eeqa
Recalling (\ref{2.k}), these formulas show 
\begin{eqnarray}
\nonumber
{d \over dx} \Phi_l(x;0) & = & 2 \sqrt{w(x)} R_k(x;0) \\
 \Phi_l(x;0) & = & - 2 {d \over dx} \Big (
\sqrt{w(x)} R_k(x;0) \Big ) \label{2.1g'}
\end{eqnarray}
in the two cases respectively.  Thus for $\beta = 1$ initial conditions 
\beqa 
\label{2.1h}
{d \over dx} S(x,y;0,\tau_y) = -2 D(x,y;0,\tau_y)
\eeqa
while for $\beta = 4$ initial conditions 
\beqa 
\label{2.1i}
\int_y^x S(x',y;0,\tau_y) \, dx' = 2 D(x,y;0,\tau_y).
\eeqa

\section{The Laguerre ensemble}
\setcounter{equation}{0}

For the Laguerre ensemble the weight function $w(x)$ and constants 
$\gamma_{k}$ are given by (\ref{2.5}), while the monic orthogonal polynomials 
and corresponding normalization have the explicit form 
\beqa
\label{3.1}
C_n(x) = n! (-1)^n L_n^a(x), \quad h_n = \Gamma(n+1) \Gamma(n+a+1).
\eeqa
This applies independent of the particular initial condition (orthogonal 
symmetry ($\beta = 1$) or symplectic symmetry ($\beta = 4$)). 
However, the skew orthogonal polynomials $R_{k}(x;\tau)$ and the quantities 
$\beta_{jk}$ depend on the initial condition. 

\subsection{$\beta = 1$ initial conditions}

In this case the monic skew orthogonal polynomials for $\tau = 0$ are known to 
be \cite{NW91}
\begin{eqnarray}
\label{3.2}
R_{2m}(x) & = & - (2m)! {d \over dx} L_{2m+1}^a (x) \: = \: (2m)!
\sum_{j=0}^{2m} {(-1)^j \over j!} C_j(x)  \nonumber \\
R_{2m+1}(x) & = & - (2m+1)!  L_{2m+1}^a (x)- (2m)! (a + 2m +1)
{d \over dx} L_{2m}^a(x) \nonumber \\
& = & C_{2m+1}(x) + (2m)!(a+2m+1) \sum_{j=0}^{2m-1}
{(-1)^j \over j!} C_j(x).
\end{eqnarray}
with corresponding normalization 
\beqa
\label{3.3}
r_n(0) = 4 \Gamma(2n+1) \Gamma(a + 2n+2).
\eeqa
These equations can readily be inverted. We find 
\begin{eqnarray}
\label{3.4}
C_{2m}(x) & = & R_{2m}(x) + (2m+a)! {m! \over (m+ a/2)!}
\sum_{j=0}^{m-1} {(j+a/2 + 1)! (2j + 2)! \over (j+1)! (2j+a+2)!} 
\nonumber \\
&& \times \Big ( {1 \over (2j+1)!} R_{2j+1}(x) -
{1 \over (2j)!} R_{2j}(x) \Big ) \nonumber \\
C_{2m+1}(x) & = & R_{2m+1}(x) + (2m+a+1)! {m! \over (m + a/2)!}
\sum_{j=0}^{m-1} {(j + a/2 + 1)! (2j + 2)! \over (j+1)! (2j + a + 2)!}
\nonumber \\
&& \times  \Big ( {1 \over (2j+1)!} R_{2j+1}(x) -
{1 \over (2j)!} R_{2j}(x) \Big )
\end{eqnarray}
(direct sustitution of (\ref{3.2}) 
into the right hand sides verifies the validity of 
these formulas). 

According to the 
definition (\ref{2.l}) with $\tau = 0$, we read off from (\ref{3.4}) that 
\begin{eqnarray}\label{3.5}
\beta_{2m \, 2m} & = & \beta_{2m+1 \, 2m+1} = 1, \qquad
\beta_{2m+1 \, 2m} = 0 \nonumber \\
\beta_{\nu \, 2j} & = & - (\nu + a)! {1 \over (2j)!}
{[(\nu/2)]! \over ([\nu/2] + a/2)!}
{(j + a/2 + 1)! \over (j+1)!} {(2j+2)! \over (2j+a+2)!} \\
\beta_{\nu \, 2j+1} & = & - {1 \over 2j + 1} \beta_{\nu \, 2j},
\end{eqnarray}
where $\nu > 2j + 1$. A most significant feature  of these formulas is that the 
dependence on $p$ and $l$ 
in $\beta_{pl}$ separates. In particular, for $n > N - 1$ and 
$N$ even we have 
\beqa
\label{3.6}
\beta_{p l} = \beta_{p \, N-1} \beta_{N-1 \, l}, \quad
0 \le l \le N-1, \: l \ne N-2.
\eeqa
Substituting this in the 
formula (\ref{2.1e}) for $S_{2}$, and making use of the final 
formula in (\ref{3.5}) in the case $(\nu, 2j + 1) = (p, N - 1)$ gives 
\begin{eqnarray}
\label{3.7}
S_2(x,y;\tau_x,\tau_y) & = & \sqrt{w(x) w(y)} \sum_{p=N}^\infty
{C_p(x) \over h_p} e^{-\gamma_p \tau_x} \beta_{p \, N-1}
\nonumber
\\ && \times \Big ( \sum_{l=0}^{N-1} \beta_{N-1 \, l} R_l(y;\tau_y)
e^{\gamma_l \tau_y} - (N-1) R_{N-2}(y;\tau_y) e^{\gamma_{N-2} \tau_y}
\Big ).
\end{eqnarray}
Now the sum 
over $l$ can be computed using (\ref{2.l}), while the sum over $p$ is 
given by (\ref{2.m}). Thus we have the explicit evaluation 
\begin{eqnarray} 
\label{3.8}
S_2(x,y;\tau_x,\tau_y) & = & \sqrt{w(x)}
\Big ( {\Phi_{N-2}(x;\tau_x) e^{\gamma_{N-2} \tau_x} \over
r_{N/2 - 1}(0)}  - \sqrt{w(x)}
{C_{N-1}(x) e^{-\gamma_{N-1} \tau_x} \over h_{N-1}} \Big )
\nonumber \\
&& \times \Big ( C_{N-1}(y) e^{\gamma_{N-1} \tau_y} -
(N - 1) R_{N-2}(y;\tau_y) e^{\gamma_{N-2} \tau_y} \Big ).
\end{eqnarray}
In  Appendix A we will 
show that in the limit $\tau_{x}, \tau_{y} \rightarrow 0$ 
this expression reduces down to the form of $S_{2}$ recently derived by 
Widom \cite{Wi98}. 

\subsubsection{The thermodynamic limit} 

In the neighbourhood 
of the spectrum edge at the origin, we will show that after 
introducing the scaled coordinates and parameters $X$ and $t$ according to 
\beqa 
x  = \frac{X}{4N} \, \quad \quad \tau = \frac{t}{2N} 
\label{3.9}
\eeqa 
(the numerical factors $1/4$
and $1/2$ are chosen for convenience), the distribution 
functions have a well defined large $N$ limit. In particular, the scaled
distribution function (\ref{2.c}), to be denoted $\rho_{(n+m)}^{\rm hard}$,
is given by
\begin{eqnarray}\label{n.1}
\lefteqn{\rho_{(n+m)}^{\rm hard}(X_1^{(1)},\dots,X_n^{(1)};t^{(1)};
X_1^{(2)},\dots,X_m^{(2)};t^{(2)})} \nonumber \\ &&
= \lim_{N \to \infty} \Big ( {1 \over 4N} \Big )^{n+m}
\rho_{(n+m)}\Big ( {X_1^{(1)}\over 4N},\dots,{X_n^{(1)}\over 4N};
{t^{(1)} \over 4N};
{X_1^{(2)} \over 4N},\dots,{X_m^{(2)}\over 4N};{t^{(2)}\over 2N}\Big )
\end{eqnarray}
{}From the formula (\ref{2.1b}) for Tdet it follows that 
$\rho_{(n+m)}^{\rm hard}$ is given by the formula (\ref{2.s}) with 
the functions $S, \tilde{S}, \tilde{I}, D$ replaced by their scaled counterparts
\begin{eqnarray}
S_1^{\rm hard}(X,Y;t_X,t_Y) & := & \lim_{N \to \infty} {1 \over 4N}
S \Big ({X \over 4N}, {Y \over 4N};{t_X \over 2N}, {t_Y \over 2N} \Big )
\label{n.2} \\
\tilde{S}_1^{\rm hard}(X,Y;t_X,t_Y) & := & \lim_{N \to \infty} {1 \over 4N}
\tilde{S} \Big ({X \over 4N}, {Y \over 4N};{t_X \over 2N}, {t_Y \over 2N} \Big )
\label{n.3} \\
\tilde{I}_1^{\rm hard}(X,Y;t_X,t_Y) & := & \lim_{N \to \infty}
\tilde{I} \Big ({X \over 4N}, {Y \over 4N};{t_X \over 2N}, {t_Y \over 2N} \Big )
\label{n.4} \\
D_1^{\rm hard}(X,Y;t_X,t_Y) & := & \lim_{N \to \infty} \Big (
{1 \over 4N} \Big )^2
D \Big ({X \over 4N}, {Y \over 4N};{t_X \over 2N}, {t_Y \over 2N} \Big )
\label{n.5}
\end{eqnarray}
where the subscripts 1 indicate $\beta = 1$ initial conditions.

Note that it was permissable to multiply ${1 \over 4N} \tilde{I}$ by $4N$
and divide ${1 \over 4N} D$ by $4N$ 
in the above definitions
because of an invariance property
of the Tdet formula (\ref{2.s}). Thus in general each $2 \times 2$ block
$Q_{jk}$ can be replaced by the $2 \times 2$ block
\begin{equation}
A^{-1} Q_{jk} A. \label{n.6'}
\end{equation}
Choosing
\begin{equation}
A = \left ( \begin{array}{cc} 1/\sqrt{4N} & 0 \\ 0 & \sqrt{4N} \end{array}
\right )
\end{equation}
then transforms  ${1 \over 4N} \tilde{I}$ and  ${1 \over 4N} D$ as
required.

\subsubsection*{Asymptotics of $S(x, y; \tau_{x}, \tau_{y})$}
Let us first 
compute the asymptotic behaviour of $\Phi_{2k-2}(x; \tau)$. This is 
required in 
the computation of $S_{2}$ (with $2k = N$), and also $\tilde{I}$. From the 
expansion (\ref{2.m}) and the explicit formulas (\ref{3.1}) and (\ref{3.5}) for the 
quantities 
$C_{\nu}(x)$, $h_{\nu}$ and $\beta_{\nu 2k-2}$ therein we have 
\begin{eqnarray}
\label{3.10}
\lefteqn{
\Phi_{2k - 2}(x;\tau) e^{\gamma_{2k-2} \tau} = \sqrt{w(x)}
r_{k-1}(0) \bigg ( - {L_{2k-1}^a (x) \over (2k - 1 + a)!} e^{-\gamma_{2k-1}
\tau}} \nonumber \\ &&
+ {(k + a/2)! (2k)! \over (2k - 1)! (2k + a)! k!}
\sum_{\nu = k}^\infty {\nu! \over (\nu + a/2)!}
\Big ( L_{2 \nu}^a(x) e^{-\gamma_{2 \nu} \tau} -
L_{2 \nu + 1}^a (x) e^{- \gamma_{2 \nu + 1} \tau} \Big ) \bigg ).
\end{eqnarray}
We now proceed to estimate the summand  for large $\nu$. 

From Stirling's formula 
\beqa
\frac{\nu !}{(\nu + a/2)!} \mathop{\sim}\limits_{\nu \to \infty}
\frac{1}{\nu^{a/2}} \,  .
\nonumber
\eeqa
For the second factor in the summand we write 
\beqa\lefteqn{
L^{a}_{2\nu}(x) e^{-\gamma_{2\nu}\tau}  - 
L^{a}_{2\nu + 1}(x) e^{-\gamma_{2\nu + 1}\tau}} \nonumber \\
&& =  
 - e^{-\gamma_{2\nu + 1}\tau} \left( 
L^{a}_{2\nu + 1}(x) - L^{a}_{2\nu}(x) - L^{a}_{2\nu}(x)
(e^{-\tau(\gamma_{2\nu} - \gamma_{2\nu + 1})} - 1) \right) \,  .
\eeqa
Using the facts that 
\beqa
e^{\tau(\gamma_{2\nu} - \gamma_{2\nu + 1})} = e^{-2\tau}\, ,   \quad  \quad  
L^{a}_{2\nu + 1}(x) - L^{a}_{2\nu}(x) = L^{a-1}_{2\nu + 1}(x) \,  ,
\label{3.11}
\eeqa
and the asymptotic formula 
\beqa
\sqrt{w(x)}\,  L_{n}^{a}(x) 
\mathop{\sim}\limits_{n \to \infty}  n^{a/2} 
J_{a}\left(2\sqrt{nx}\right) 
\label{3.12}
\eeqa
where $J_{a}(z)$ denotes the Bessel function,  we see that 
\beqa
\lefteqn{\left. \sqrt{w(x)}\, \frac{\nu !}{(\nu + a/2)!} 
 \left(L_{2\nu}^{a}(x) e^{-\gamma_{2\nu}\tau}  - 
L^{a}_{2\nu + 1}(x) e^{-\gamma_{2\nu + 1} \tau} \right)
\right|_{x = X/4N \atop \tau = t/2N}} \nonumber \\
 & \sim  &  -\nu^{-a/2} e^{-(2\nu /N)t} 
\bigg( \sqrt{\frac{X}{4N}}(2\nu)^{(a-1)/2} 
J_{a-1}\Big(\sqrt{\frac{2jX}{N}}\Big) + \frac{t}{N} (2\nu)^{a/2} 
J_{a}\Big(\sqrt{\frac{2jX}{N}}\Big)\bigg) 
\nonumber \\
& =  & -\frac{2^{a/2}}{N}u^{-a/2}\frac{d}{du}
\left(e^{-ut} u^{a/2} J_{a}\left(\sqrt{uX}\right) \right), \quad \quad 
u := 2\nu /N \,  .
\label{3.13}
\eeqa
(The equality is obtained by making use of a suitable differentiation formula for 
the Bessel function). 

Substituting (\ref{3.13}) 
in (\ref{3.10}) gives the leading order behaviour of the sum 
over $\nu$ as a Riemann integral.  Furthermore, use of (\ref{3.12}) gives
\beqa
\left. \sqrt{w(x)} \, \frac{L^{a}_{2k - 1}(x)}{(2k - 1 + a)!}\,
 e^{-\gamma_{2k-1}\tau}\right|_{x = X/4N \atop \tau = t/2N}
\sim  
\frac{N^{-a/2}}{(2k - 1)!} \, s^{-a/2} J_{a}\left(\sqrt{sX}\right)\, e^{-st},
\label{3.14} 
\eeqa
$s := 2k/N$, so in total we have 
\beqa
\Phi_{2k-2}\left(\frac{X}{4N}; \frac{t}{2N}\right) e^{\gamma_{2k-2}t/2N} & \sim &
r_{k-1}(0) \frac{N^{-a/2}}{(2k - 1)!} 
\bigg ( \frac{1}{2}\left( 1 - 2s^{-a/2}\right) J_{a}\left(\sqrt{sX}\right) 
e^{-st}  \nonumber \\
&  &   -  
\frac{a}{4} \int_{s}^{\infty} \frac{e^{-ut}}{u} J_{a}\left(\sqrt{uX}\right) 
du \bigg )  \,  .
\label{3.15}
\eeqa
The expansions (\ref{3.14}) and (\ref{3.15}) with $2k = N$ and thus $s = 1$ 
suffice to 
specify the $x$-dependent factor in (\ref{3.8}). Thus we have 
\beqa
\lefteqn{\left. 
\left(\frac{\Phi_{N-2}(x; \tau_{x})}{r_{N/2 - 1}(0)}  -  \sqrt{w(x)}\, 
\frac{C_{N-1}(x)e^{-\gamma_{N-1} \tau_{x}}}{h_{N-1}}\right) 
\right|_{x = X/4N \atop \tau_{x} = t_{X}/2N}}
\nonumber \\ 
& & \sim  \frac{N^{-a/2}}{2(N-1)!} \left( 
J_{a}\left(\sqrt{X}\right) e^{-t_{X}} - \frac{a}{2} \int_{1}^{\infty} 
\frac{e^{-ut_{X}}}{u} J_{a} \left( \sqrt{uX}\right) \, du \right)  \,  . 
\label{3.16}
\eeqa

Next we will compute the asymptotic form of $R_{2j}(y; \tau_{y})$. In the case 
$2j = N - 2$ this is required for the $y$-dependent factor in (\ref{3.8}). From 
(\ref{2.i}), (\ref{3.2}) and (\ref{3.1}) we see that 
\beqa
R_{2j}(y; \tau )\, e^{\gamma_{2j} \tau} = (2j)! \sum_{l=0}^{2j}L_{l}^{a}(y)\, 
e^{\gamma_{l}\tau} \,  . 
\label{3.16'}
\eeqa
The asymptotic form of this expression is deduced by using (\ref{3.12}) to 
approximate the summand, which leads to the result 
\beqa
\left. \sqrt{w(y)} R_{2j}(y; \tau_{y}) e^{\gamma_{2j} \tau_{y}}
\right|_{y = Y/4N \atop \tau_{y} = t_{Y}/2N}
\sim (2j)! \, 
N^{1+a/2} \int_{0}^{s}v^{a/2}J_{a}\left(\sqrt{vY}\right)\, e^{vt_{Y}}\, dv 
\,  , 
\label{3.17}
\eeqa
where here $s = 2j/N$. Setting $s = 1$ and making further use of (\ref{3.14}) 
we see that 
\beqa
\lefteqn{\sqrt{w(y)} \left(C_{N-1}(y)e^{\gamma_{N-1}\tau_{y}} - 
(N-1) R_{N-2}(y; \tau_{y}) e^{\gamma_{N-2} \tau_{y}}\right)}
\nonumber \\ & & \hspace*{2cm} \sim  -(N-1)! N^{1+a/2} 
\int_{0}^{1}v^{a/2}J_{a}\left(\sqrt{vY}\right)\, e^{vt_{Y}}\, dv \,   .
\label{3.18}
\eeqa
Combining (\ref{3.16}) and (\ref{3.18}) in the formula (\ref{3.8}) for $S_{2}$ gives 
\beqa
S_{2}\left(\frac{X}{4N}, \frac{Y}{4N}; \frac{t_{X}}{2N}, \frac{t_{Y}}{2N} 
\right) 
& \sim  & 
\frac{N}{2}\left( - e^{-t_{X}} J_{a}\left(\sqrt{X}\right) + \frac{a}{2} 
\int_{1}^{\infty} \frac{e^{-ut_{X}}}{u}J_{a}\left(\sqrt{uX}\right)\, du\right) 
\nonumber \\  &  &  \times 
\left( \int_{0}^{1}v^{a/2}J_{a}\left(\sqrt{vY}\right)\, e^{vt_{Y}}\, 
dv \right)\,  .
\label{3.19}
\eeqa

The 
asymptotics of $S_{1}$ can be determined by using (\ref{3.12}) and Stirling's 
formula to approximate the integrand. This gives 
\beqa
S_{1}\left(\frac{X}{4N}, \frac{Y}{4N}; \frac{t_{X}}{2N}, \frac{t_{Y}}{2N} 
\right) 
\sim  N \int_{0}^{1} J_{a}\left(\sqrt{uX}\right)\,J_{a}\left(\sqrt{uY}\right)\, 
e^{-u(t_{X} - t_{Y})}\, du \,  . 
\label{3.20}
\eeqa
Since $S$ is defined as the sum of $S_{1}$ and $S_{2}$, its asymptotic form 
is given 
by the sum of (\ref{3.19}) and (\ref{3.20}). Substituting into (\ref{n.2})
gives
\begin{eqnarray}\label{3.20'} \lefteqn{
S_1^{\rm hard}(X,Y;t_X,t_Y) = {1 \over 4} \int_0^1 J_a(\sqrt{uX})
J_a(\sqrt{uY}) e^{-u(t_X - t_Y)} \, du} \nonumber \\ &&
+ {1 \over 8} \Big ( - e^{-t_X} J_a(\sqrt{X}) + {a \over 2}
\int_1^\infty {e^{-u t_X} \over u} J_a(\sqrt{uX}) \, du \Big )
\int_0^1 dv \, J_a(\sqrt{vY}) e^{t_Y v} v^{a/2}.
\end{eqnarray}

\subsubsection*{Asymptotics of $\tilde{I} (x, y; \tau_{x}, \tau_{y})$} 

So as to induce a cancellation at leading order, we first perform some minor 
manipulation to the definition (\ref{2.p}), by way of subtracting and adding 
$(2l + 1) e^{\gamma_{2l} (\tau_{x} + \tau_{y})} \Phi_{2l}(x; \tau_{x}) 
\Phi_{2l}(y; \tau_{y})$ to the summand, so that it reads 
\beqa
\tilde{I} (x, y; \tau_{x}, \tau_{y}) & = & \sum_{l = N/2}^{\infty} 
\frac{1}{r_{l}(0)} \left\{ e^{\gamma_{2l}\tau_{x}} \Phi_{2l}(x; \tau_{x})
\left( e^{\gamma_{2l+1}\tau_{y}} \Phi_{2l+1}(y; \tau_{y}) \right. \right. 
\nonumber \\
&  &  \left. \left. - (2l + 1) e^{\gamma_{2l}\tau_{y}} \Phi_{2l}(y; \tau_{y}) \right) - 
(x \leftrightarrow y) \right\} \,  . 
\label{3.21}
\eeqa
The asymptotics 
of $e^{\gamma_{2l} \tau_{x}} \Phi_{2l}(x; \tau_{x})$ can be read 
off from (\ref{3.15}). For the $y$-dependent factor of the first term of the 
summand we note from the definitions (\ref{2.m}) and the explicit formulas 
(\ref{3.1}) and (\ref{3.5}) that 
\beqa
\lefteqn{e^{\gamma_{2l+1}\tau} \Phi_{2l+1}(y; \tau) 
- (2l + 1) e^{\gamma_{2l}\tau} \Phi_{2l}(y; \tau)} \nonumber \\ 
& & = \sqrt{w(y)}\,  r_{l}(0) \left( (2l+1)
\frac{L_{2l+1}^{a}(y)e^{-\gamma_{2l+1}\tau}}{(2l+1+a)!} - 
\frac{L_{2l}^{a}(y)e^{-\gamma_{2l}\tau}}{(2l+a)!}\right) \,  . 
\nonumber
\eeqa
Thus, using (\ref{3.3}) and the asymptotic formula (\ref{3.12}) we have 
\beqa
 \lefteqn{\Big( e^{\gamma_{2l+1}\tau} \Phi_{2l+1}(y; \tau) 
- (2l + 1) e^{\gamma_{2l}\tau} \Phi_{2l}(y; \tau) \Big)
\bigg|_{y = Y/4N \atop \tau = t/2N}} \nonumber \\
&  &  \sim 4(2l+1)! N^{-1 + a/2} \left\{ 
\frac{d}{ds}\left(e^{-st}J_{a}\left(\sqrt{sY}\right)\right) - 
as^{-1+a/2}J_{a}\left(\sqrt{sY}\right)e^{-st}\right\}\,  , \quad  s = 2l/N \,  . 
\nonumber
\eeqa
Substituting this result, together with (\ref{3.15}) (after replacing $k$ by $l + 1$) in 
(\ref{3.21}) shows that 
\beqa
\tilde{I}_1^{\rm hard}(X,Y;t_X,t_Y) =
 \int_{1}^{\infty}  ds \left\{ \left(1- \frac{2}{s^{a/2}}\right)
J_{a}\left(\sqrt{sX}\right) e^{-st_{X}} - \frac{a}{4}\int_{s}^{\infty} 
\frac{e^{-ut_{X}}}{u} J_{a}\left(\sqrt{uX}\right)\,du \right\} 
\nonumber \\   \times
\left\{ \frac{d}{ds}\left(e^{-st_{y}}J_{a}\left(\sqrt{sY}\right)\right) - 
as^{-1+a/2}J_{a}\left(\sqrt{sY}\right)
e^{-st_{Y}}\right\} - (X \leftrightarrow Y) \, .
\label{3.22}
\eeqa

\subsubsection*{Asymptotics of $g(x, y; \tau)$}

According to (\ref{2.3}) and (\ref{2.5}) 
\beqa
g(x, y; \tau) = (xy)^{a/2}e^{-(x+y)/2} \sum_{j=0}^{\infty} \frac{j!}{(j+a)!} 
L_{j}^{a}(x)L_{j}^{a}(y) e^{-2(j+(a+1)/2)\tau} \, . 
\label{3.22a}
\eeqa
This summation can be evaluated in closed form \cite{Er}, giving
\beqa
g(x, y; \tau) = e^{-(x+y)/2}e^{-(2a+1)\tau}\left(1 - e^{-2\tau}\right)^{-1} 
\exp \left( -\frac{e^{-2\tau}}{1 - e^{-2\tau}}(x+y) \right) 
I_{a}\left(2\frac{\sqrt{xy}\, e^{-\tau}}{1 - e^{-2\tau}}\right)\, ,
\nonumber
\eeqa
where $I_a$ denotes the Bessel function of pure imaginary argument.
Consequently 
\beqa
g\left( \frac{X}{4N}, \frac{Y}{4N}; \frac{t_{X} - t_{Y}}{2N}\right) \sim 
\frac{N}{t_{X}-t_{Y}} 
\exp \left( -\frac{X+Y}{t_{X} - t_{Y}} \right) 
I_{a}\left(\frac{\sqrt{XY}}{2(t_{X} - t_{Y})}\right) \,  .
\label{3.22'}
\eeqa

Alternatively, proceeding as in the derivation of (\ref{3.20}) we can deduce
from (\ref{3.22a}) that
\begin{equation}\label{m.1}
g\Big ({X \over 4N}, {Y \over 4N}; {t_X - t_Y \over 2N} \Big ) \sim
N \int_0^\infty J_a(\sqrt{uX})  J_a(\sqrt{uY}) e^{-u (t_X - t_Y)} \, du.
\end{equation}
The immediate 
merit of this representation is that when substituted in (\ref{2.r}), partial
cancellation occurs with (\ref{3.20}). Adding the result to (\ref{3.19})
and substituting in (\ref{n.3}) gives
\begin{eqnarray}\label{m.2} \lefteqn{
\tilde{S}_1^{\rm hard}(X,Y;t_X,t_Y) = - {1 \over 4} \int_1^\infty J_a(\sqrt{uX})
J_a(\sqrt{uY}) e^{-u(t_X - t_Y)} \, du} \nonumber \\ &&
+ {1 \over 8} \Big ( - e^{-t_X} J_a(\sqrt{X}) + {a \over 2}
\int_1^\infty {e^{-u t_X} \over u} J_a(\sqrt{uX}) \, du \Big )
+ \int_0^1 dv \, J_a(\sqrt{vY}) e^{t_Y v} v^{a/2}.
\end{eqnarray}

\subsubsection*{Asymptotics of $D(x, y; \tau_{x}, \tau_{y})$}

According to (\ref{2.q}) $D$ is defined in terms of $\{R_{j}(x; \tau)\}$. Now 
$R_{2j}(y; \tau)$ is specified by (\ref{3.16'}). Deriving the analogous formula for 
$R_{2j+1}(x; \tau)$, which follows from (\ref{3.2}), (\ref{2.i}) and (\ref{3.1}), we see 
that
\beqa
R_{2j+1}(x; \tau) e^{\gamma_{2j+1}\tau}
& =  &  -(2j+1)! L_{2j+1}^{a}(x) e^{\gamma_{2j+1}\tau} - 
(2j)! (2j+a + 1) L_{2j}^{a}(x) e^{\gamma_{2j}\tau} \nonumber \\  & & 
\quad \quad + (2j+a + 1) R_{2j}(x; \tau) e^{\gamma_{2j}\tau} \,  .
\nonumber
\eeqa
Hence
\beqa
\lefteqn{\left( e^{\gamma_{2j+1}\tau_{x}} R_{2j+1}(x; \tau_{x}) 
e^{\gamma_{2j}\tau_{y}} R_{2j}(y; \tau_{y}) - (x \leftrightarrow y) \right) =} 
\nonumber \\ 
& & \left(-(2j+1)!  L_{2j+1}^{a}(x) e^{\gamma_{2j+1}\tau_{x}} - 
(2j)! (2j+a + 1) L_{2j}^{a}(x) e^{\gamma_{2j}\tau_{x}}\right) 
R_{2j}(y; \tau_{y}) e^{\gamma_{2j}\tau_{y}}  - (x \leftrightarrow y)  \,  .
\nonumber
\eeqa
Now the asymptotic form 
of $R_{2j}(y; \tau_{y})$ is specified by (\ref{3.17}) while we 
see from (\ref{3.12}) that
\beqa
\lefteqn{\sqrt{w(x)} \left(-(2j+1)!  
L_{2j+1}^{a}(x) e^{\gamma_{2j+1}\tau_{x}} - 
(2j)! (2j+a + 1) L_{2j}^{a}(x) e^{\gamma_{2j}\tau_{x}}\right)  
\bigg|_{x = X/4N \atop \tau_{x} = t_{X}/2N}} \nonumber \\ & & \quad \quad \quad 
\sim -2 (2j+1)! e^{st_{X}} (Ns)^{a/2} J_{a}\left( \sqrt{sX}\right) \, , \quad s = 2j/N \,  .
\nonumber
\eeqa
Making use also of the explicit form (\ref{3.3}) of $r_{n}(0)$ 
and using the definition (\ref{n.5}) we thus find 
\beqa\lefteqn{
D_1^{\rm hard}(X,Y;t_X,t_Y) = - 4^{-3}  } \nonumber \\ && \times
\left\{ \int_{0}^{1}ds\, e^{st_{X}} s^{-a/2}  
J_{a}\left( \sqrt{sX}\right)  
\int_{0}^{s}dv \, e^{vt_{Y}}v^{a/2}J_{a}\left( \sqrt{vY}\right) 
- (X \leftrightarrow Y) \right\} \,  .
\label{3.23}
\eeqa

\subsection{$\beta = 4$ initial conditions}

The method of evaluation of the 
quantities $S, \, \tilde{I}$, and $D$ for the Laguerre 
ensemble in the 
case of initial conditions with symplectic symmetry ($\beta = 4$) 
closely 
parallels that just given for $\beta = 1$ initial conditions. First, we note 
from previous 
work \cite{NW91} that the monic skew orthogonal polynomials at $\tau = 0$ 
have the explicit form 
\beqa
R_{2j}(x) &=& 2^{2j}j!\left(j + (a-1)/2\right)! \sum_{l=0}^{j} 
\frac{(2l)!}{l! \, 2^{2l}} \frac{1}{\left(l + (a-1)/2\right)!} L_{2l}^{a-1}(x) 
\nonumber \\
& = &  2^{2j}j!\left(j + (a-1)/2\right)! \sum_{l=0}^{j} 
\frac{1}{l! \, 2^{2l} \left(l + (a-1)/2\right)!} \left(C_{2l}(x) + 2lC_{2l-1}(x) \right) \,  , 
\nonumber \\
R_{2j+1}(x) &=& -(2j+1)! \,   L_{2j+1}^{a-1}(x) 
\nonumber \\
& = &  C_{2j+1}(x) + (2j+1)C_{2j}(x)  \,   , 
\label{3.24}
\eeqa
with corresponding normalization
\beqa
r_{n}(0) = \Gamma(2n+2)\, \Gamma(a + 2n + 1) \,   . 
\label{3.25}
\eeqa

The inversion of these equations gives that the quantities 
$\beta_{n\, j}$ in (\ref{2.l}) 
have the explicit values 
\beqa
\beta_{2m\, 2m} &=& \beta_{2m+1\, 2m+1} = 1 \, , \quad \quad  
\beta_{2m+1\, 2m} = - (2m+1) \,   ,   \nonumber \\ 
\beta_{\nu\, 2j+1} &=& (-1)^{\nu + 1} \frac{\nu !}{(2j+1)!} \,  ,  \quad \quad 
\beta_{\nu\, 2j} = a \,  \beta_{\nu\, 2j+1} \,   , 
\label{3.26}
\eeqa
for $\nu > 2j + 1$. Thus 
the factorization formula (\ref{3.6}) again holds, which when 
substituted in (\ref{2.1e}) and after making use of the explicit value of 
$\beta_{N-1\, N-2}$ and $\beta_{p\, N-1}$, and (\ref{2.l}), implies 
\beqa
S_{2}(x, y; \tau_{x}, \tau_{y}) & = & \sqrt{w(x)w(y)} \sum_{p = N}^{\infty} 
\frac{C_{p}(x)}{h_{p}}\, e^{-\gamma_{p}\tau_{x}}\, \beta_{p\, N-1} 
\left\{ C_{N-1}(y) e^{\gamma_{N-1}\tau_{y}} \right. \nonumber \\
& & \quad \left. + (a + N-1) R_{N-2}(y; \tau_{y}) 
e^{\gamma_{N-2}\tau_{y}} \right\} \,  . 
\nonumber
\eeqa
The sum over $p$ is, 
after minor manipulation, formally identical to the sum over $p$ 
in (\ref{3.7}), 
so it can be summed using (\ref{2.m}).  Thus $S_{2}$ has the explicit 
form 
\beqa
S_{2}(x, y; \tau_{x}, \tau_{y}) & = & \sqrt{w(y)} \left( 
\frac{\Phi_{N-2}(x; \tau_{x})}{r_{N/2 -1}(0)} e^{\gamma_{N-2}\tau_{x}} - 
\sqrt{w(x)} \frac{C_{N-1}(x)}{h_{N-1}} e^{-\gamma_{N-1}\tau_{x}} \right) 
\nonumber \\
& & \times 
\left(C_{N-1}(y) e^{\gamma_{N-1}\tau_{y}} + (a + N-1) R_{N-2}(y; \tau_{y}) 
e^{\gamma_{N-2}\tau_{y}} \right) \,  . 
\label{3.27}
\eeqa
As with (\ref{3.8}), 
in the limit $\tau_{x}, \tau_{y} \rightarrow 0$ this reduces down to 
the form of $S_{2}$ 
derived by Widom \cite{Wi98} in the static $\beta = 4$ theory. 
The details of the verification are given in Appendix A. 

\subsubsection{The thermodynamic limit}
Although the scaled distribution function is still given by (\ref{n.1}),
the required scaling of the elements of the quaternion determinant
(\ref{2.s}) differs slightly from the formulas (\ref{n.2})--(\ref{n.5})
appropriate with $\beta = 1$ initial conditions. In particular (\ref{n.4})
and (\ref{n.5}) should now read
\begin{eqnarray}
\tilde{I}_4^{\rm hard}(X,Y;t_X,t_Y) & := & \lim_{N \to \infty}
\Big ( {1 \over 4N} \Big )^2
\tilde{I} \Big ({X \over 4N}, {Y \over 4N};{t_X \over 2N}, {t_Y \over 2N} \Big )
\label{n.6} \\
D_4^{\rm hard}(X,Y;t_X,t_Y) & := & \lim_{N \to \infty} 
D \Big ({X \over 4N}, {Y \over 4N};{t_X \over 2N}, {t_Y \over 2N} \Big )
\label{n.7}
\end{eqnarray}
while (\ref{n.2}) and (\ref{n.3}) remain the same, except that the subscripts
1 on the right hand sides should be replaced by 4.

\subsubsection*{Asymptotics of $S(x, y; \tau_{x}, \tau_{y})$}

From the expansion 
(\ref{2.m}) and the explicit formulas (\ref{3.1}) and (\ref{3.26}) we have 
\beqa
\Phi_{2k-2}(x; \tau)e^{\gamma_{2k-2}\tau} 
= -\frac{r_{k-1}(0) \sqrt{w(x)}}{(2k-1)!} 
\sum_{\nu = 2k-1}^{\infty} \frac{L_{\nu}^{a}(x)\, \nu !}{(\nu + a)!} 
\, e^{-\gamma_{\nu}\tau}\, . 
\nonumber
\eeqa
Applying the 
asymptotic expansion (\ref{3.12})  and Stirling's formula shows that 
\beqa
\Phi_{2k-2}\left(\frac{X}{4N}; \frac{t}{2N}\right)
e^{\gamma_{2k-2} t/2N}  \sim -\frac{r_{k-1}(0)}{(2k-1)!} N^{1 - a/2} 
\int_{u}^{\infty} dr \, r^{-a/2} e^{-rt} J_{a}\left(\sqrt{rX}\right)  \, , 
\label{3.28}
\eeqa
where $u := 2k/N$.  
Setting $u = 1$ gives 
the asymptotic form of $\Phi_{N-2}(x; \tau_{x})/r_{N/2 -1}(0)$ 
as required in (\ref{3.27}), 
and comparison with (\ref{3.14}) shows that this term 
dominates the one involving 
$C_{N-1}(x)$. Thus here the formula (\ref{3.16}) applies 
with the right hand side replaced by 
\beqa
- \frac{N^{1 - a/2}}{(N-1)!}  
\int_{u}^{\infty} dr \, r^{-a/2} e^{-rt} J_{a}\left(\sqrt{rX}\right) \,   .
\label{3.29}
\eeqa
For the $y$-dependent factor 
in (\ref{3.27}), we note from (\ref{2.i}), (\ref{3.1}), and 
(\ref{3.16'}) that 
\beqa
\lefteqn{R_{2j}(y; \tau)  e^{\gamma_{2j}\tau} 
=} \nonumber \\
& &  2^{2j}j!  \left( j + (a-1)/2\right)! \sum_{l=0}^{j} 
\frac{(2l)!}{l!\, 2^{2l}} \frac{1}{(l + (a-1)/2)!} 
\left( L_{2l}^{a}(y) e^{\gamma_{2l}\tau} - L_{2l-1}^{a}(y) e^{\gamma_{2l-1}\tau}\right)  .
\nonumber 
\eeqa
Use of (\ref{3.12}) 
and Stirling's formula gives for the corresponding asymptotic form 
\beqa
\lefteqn{\left. \frac{1}{(2j+1)!}\sqrt{w(y)} R_{2j}(y; \tau) e^{\gamma_{2j}\tau}
\right|_{y = Y/4N \atop \tau = t/2N}} \nonumber \\
& \sim & \frac{N^{a/2-1}}{2} s^{-1 + a/2} \int_{0}^{s} 
u^{-a/2} \frac{d}{du} \left( e^{ut} u^{a/2} J_{a}\left(\sqrt{uY}\right) \right) du \, , 
\quad s := 2j/N\,  . 
\label{3.30}
\eeqa
Adding this with $s = 1$ to the asymptotic form of 
$\sqrt{w(y)} 
C_{N-1}(y) e^{\gamma_{N-1} \tau_{y}}/(N-1)!$ deduced from (\ref{3.14}), 
and multiplying the result by (\ref{3.29}) gives that 
\beqa
\lefteqn{S_{2}\left(\frac{X}{4N}, 
\frac{Y}{4N}; \frac{t_{X}}{2N}, \frac{t_{Y}}{2N} \right) 
\sim  -N\int_{1}^{\infty} dr \, r^{-a/2} e^{-rt_{X}}J_{a}\left(\sqrt{rX}\right)}
\nonumber \\ &  & \times
\left( - J_{a}\left(\sqrt{Y}\right) e^{t_{Y}} + 
\frac{1}{2} \int_{0}^{1}u^{-a/2} \frac{d}{du}\left( e^{u t_{Y}} u^{a/2} 
J_{a}\left(\sqrt{uY}\right)\right) \,du \right)\,  .
\label{3.31}
\eeqa

The asymptotic form of $S_{1}$ can be read off from (\ref{3.20}) 
since according to its 
definition (\ref{2.1d}), 
its value is the same for $\beta = 1$ and 4.  The asymptotic form of 
$S$ itself is thus given by the sum of (\ref{3.20}) and (\ref{3.31}).  
Thus we have
\begin{eqnarray}\lefteqn{
S_4^{\rm hard}(X,Y;t_X,t_Y) = } \nonumber \\ &  & {1 \over 4} \int_0^1
J_a(\sqrt{uX}) J_a(\sqrt{uY}) e^{-u(t_X - t_Y)} \, du
- {1 \over 4} \int_1^\infty dr \, r^{-a/2} e^{-r t_X}
J_a(\sqrt{rX}) \nonumber \\ &&
\times \bigg ( - J_a(\sqrt{Y}) e^{t_Y} + {1 \over 2}
\int_0^1 u^{-a/2} {d \over du} \Big ( e^{ut_Y} u^{a/2}
J_a(\sqrt{uY}) \Big ) du \bigg ) \label{m.3}
\end{eqnarray}

\subsubsection*{Asymptotics of $\tilde{S}(x, y; \tau_{x}, \tau_{y})$}
The single particle Green function $g$ as specified by (\ref{2.3})
is independent of the initial condition, so me can use the formula
(\ref{m.1}). Dividing by $4N$ and subtracting from (\ref{m.3}) shows
\begin{eqnarray}\lefteqn{
\tilde{S}_4^{\rm hard}(X,Y;t_X,t_Y)} 
\nonumber \\ && =  - {1 \over 4} \int_1^\infty
J_a(\sqrt{uX}) J_a(\sqrt{uY}) e^{-u(t_X - t_Y)} \, du
- {1 \over 4} \int_1^\infty dr \, r^{-a/2} e^{-r t_X}
J_a(\sqrt{rX}) \nonumber \\ &&
\times \bigg ( - J_a(\sqrt{Y}) e^{t_Y} + {1 \over 2}
\int_0^1 u^{-a/2} {d \over du} \Big ( e^{ut_Y} u^{a/2}
J_a(\sqrt{uY}) \Big ) du \bigg ) \label{m.4}
\end{eqnarray}
\subsubsection*{Asymptotics of $\tilde{I}(x, y; \tau_{x}, \tau_{y})$}

The definition (\ref{2.p}) gives that $\tilde{I}$ is defined in terms of $\{\Phi_{l}(x; \tau)\}$. 
The asymptotic form of $\Phi_{2k-2}(x; \tau)$ is given by (\ref{3.28}). Instead of 
deriving an analogous 
expression for $\Phi_{2k-1}(y; \tau)$, we note from (\ref{2.m}), 
(\ref{3.26}), (\ref{3.1}) and the formula (\ref{3.28}) for $\Phi_{2k-2}$ that 
\beqa
\lefteqn{\Phi_{2k-1}(y; \tau)e^{\gamma_{2k-1}\tau} =} \nonumber \\ 
& - & \frac{\sqrt{w(y)}\, r_{k-1}(0)}{(2k-2 + a)!} 
\left( L_{2k-2}^{a}(y) 
e^{-\gamma_{2k-2}\tau} + L_{2k-1}^{a}(y) e^{-\gamma_{2k-1}\tau}
\right) - a \Phi_{2k-2}(y; \tau) \,  . 
\label{3.32}
\eeqa
Thus
\beqa
\lefteqn{e^{\gamma_{2l}\tau_{x}}\Phi_{2l}(x; \tau_{x})
e^{\gamma_{2l+1}\tau_{y}}\Phi_{2l+1}(y; \tau_{y}) - (x \leftrightarrow y) 
=}  \nonumber \\ 
&  &  e^{\gamma_{2l}\tau_{x}}\Phi_{2l}(x; \tau_{x})
\left(-\frac{\sqrt{w(y)}\, r_{l}(0)}{(2l + a)!} \right)
\left( L_{2l}^{a}(y) 
e^{-\gamma_{2l}\tau_{y}} + L_{2l+1}^{a}(y) e^{-\gamma_{2l+1}\tau_{y}}
\right) - ( x \leftrightarrow y)   .  \nonumber \\
\label{3.33}
\eeqa
According to (\ref{3.12}) and (\ref{3.25}), the product of the second and third 
factors in the first term have the same leading asymptotic behaviour
\beqa
-(2l+1)! \, 
4(vN)^{a/2} J_{a}\left(\sqrt{vY}\right)\, 
e^{-v\tau} \,  , \quad  v = 2l/N \, .  
\label{3.34}
\eeqa
Multiplying this by 
(\ref{3.28}) (with $k \mapsto l + 1$) and substituting in (\ref{2.p}) 
gives that the desired asymptotic form is 
\beqa
\lefteqn{I_4^{\rm hard}(X,Y;t_X,t_Y)} \nonumber  \\
&& = 
{1 \over 4} \int_{1}^{\infty}\left\{
\int_{v}^{\infty}dr \, 
r^{-a/2}  e^{-r t_{X}} J_{a}\left(\sqrt{rX}\right)\right\}
v^{a/2} J_{a}\left(\sqrt{vY}\right) e^{-vt_{Y}} \, dv - (X \leftrightarrow Y)  .
\label{3.35}
\eeqa

\subsubsection*{Asymptotics of $D(x, y; \tau_{x}, \tau_{y})$}

Here we require the asymptotic form of $\{R_{l}(x; \tau)\}$. For $l$ even this 
is given by 
(\ref{3.30}). 
On the other hand, for $l$ odd we see from (\ref{2.i}), (\ref{3.1}) and 
(\ref{3.24}) that 
\beqa
R_{2j+1}(y; \tau)e^{\gamma_{2j+1}\tau} = -(2j+1)! 
\left( L_{2j+1}^{a}(y) 
e^{\gamma_{2j+1}\tau} - L_{2j}^{a}(y) e^{\gamma_{2j}\tau}
\right)  \,  . 
\label{3.36}
\eeqa
Use of (\ref{3.12}) shows 
\beqa
\left. \sqrt{w(y)} \,  R_{2j+1}(y; \tau) e^{\gamma_{2j+1}\tau}
\right|_{y = Y/4N \atop \tau = t/2N} 
\sim -(2j+1)! N^{a/2 - 1} \frac{d}{ds}\left( 
e^{st} s^{a/2} J_{a}\left(\sqrt{sY}\right) \right)  \, , 
\label{3.37}
\eeqa
$s := 2j/N$. Multiplying this by $(2j + 1)!$ times (\ref{3.30}) (with $Y$ replaced by $X$) 
and dividing by (\ref{3.25}) we thus have 
\beqa
D_4^{\rm hard}(X,Y;t_X,t_Y)
& \sim  &  -\frac{1}{4} \int_{0}^{1}ds\, s^{-a/2} \left\{
\int_{0}^{s} u^{-a/2}  \frac{d}{du}\left( 
e^{u t_{X}} u^{a/2} J_{a}\left(\sqrt{uX}\right) \right) du \right\} 
\nonumber \\  &  & \quad \times 
\frac{d}{ds}\left( e^{st_{Y}} s^{a/2} J_{a}\left(\sqrt{sY}\right) \right) 
 - (X \leftrightarrow Y) \,  .
\label{3.38}
\eeqa 
%%%%%%%%%%%%%%%%%%%%%%%%%%%%%%%%%%%%%%%%%%%%%%%%%%%%%%%%%%%

\section{Gaussian ensemble}
\setcounter{equation}{0}

For the Gaussian ensemble, we see from (\ref{2.4}) that the corresponding monic 
orthogonal polynomials and normalization are 
\beqa
C_{n}(x) = 2^{-n}\, H_{n}(y), \quad \quad h_{n} = 2^{-n}\, \pi^{1/4}\, n! \, .
\label{4.1}
\eeqa
The skew orthogonal 
polynomials for both $\beta = 1$ and $\beta = 4$ initial conditions 
at $\tau = 0$ are known in terms of the $C_{n}(x)$ (see e.g.~\cite{MM91}).  
Using these 
expansions, the general formula (\ref{2.i}) then gives 
\begin{eqnarray}
R_{2m}(x;\tau) e^{\gamma_{2m} \tau} & = & C_{2m}(x) e^{\gamma_{2m} \tau} 
\noindent \\
R_{2m+1}(x;\tau) e^{\gamma_{2m+1} \tau} & = & 
C_{2m+1}(x) e^{\gamma_{2m+1} \tau} - m C_{2m-1}(x)  e^{\gamma_{2m-1} \tau}
\noindent\\
r_{m}(0) & = & 2^{-2m+1} \sqrt{\pi} \Gamma(2m+1)
\label{4.2}
\end{eqnarray}
for $\beta = 1$ initial conditions, and 
\begin{eqnarray}
R_{2m}(x;\tau) e^{\gamma_{2m} \tau} & = & m! 
\sum_{l=0}^m e^{\gamma_{2l} \tau} {C_{2l}(x) \over l!} \nonumber \\
R_{2m+1}(x;\tau) e^{\gamma_{2m+1} \tau} & = & C_{2m+1}(x) e^{\gamma_{2m+1}
\tau} \\
r_m(0) & =  & 2^{-2m} \sqrt{\pi} \Gamma(2m+2)
\label{4.3}
\end{eqnarray}
for $\beta = 4$ initial conditions. Inverting these formulas for $\{R_{j}(x)\}$ in terms of 
$\{C_{l}(x)\}$ according to (\ref{2.l}) gives 
\beqa
\label{4.4}
\beta_{2m \, j} = \delta_{2m,j}, \qquad
\beta_{2m+1 \, j} = \left \{
\begin{array}{ll}m!/l!, & j=2l+1, \: l=0,1,\dots,m \\
0, & j=2l, \: l=0,1,\dots,m \end{array} \right.
\eeqa
for $\beta = 1$, while for $\beta =4$ we find 
\beqa
\label{4.5}
\beta_{2m \, j} = \left \{ \begin{array}{lll}1, & j=2m \\
-m, & j=2m-2 \\ 0, & {\rm otherwise} \end{array} \right. \qquad
\beta_{2m+1 \, j} = \delta_{2m+1, j}.
\eeqa

Here we will use these explicit 
values in the general formulas of Section 2 to compute the 
matrix elements in (\ref{2.s}). 
In particular, we will give their asymptotic form 
in the 
neighbourhood of the soft edge, which is specified by introducing the scaled 
variables $X$ and $t$ according to 
\beqa
x = \sqrt{2N} + 
\frac{X}{2^{1/2}N^{1/6}}\, , \quad \quad \tau = \frac{t}{N^{1/3}}\,   .
\label{4.6}
\eeqa
The corresponding scaled distribution function (\ref{2.c}), to be denoted
$\rho_{(n+m)}^{\rm soft}$, is given by
\begin{eqnarray}\lefteqn{\rho_{(n+m)}^{\rm soft}(X_1^{(1)},\dots,X_n^{(1)};
t^{(1)};X_1^{(2)},\dots,X_m^{(2)};t^{(2)})} \nonumber \\ &&
= \lim_{N \to \infty}
\Big ( {1 \over 2^{1/2} N^{1/6}} \Big )^{n+m} \rho_{(n+m)}\Big ( \sqrt{2N} +
 {X_1^{(1)} \over 2^{1/2} N^{1/6}}, \dots,
\label{r.1}\nonumber \\ &&
\sqrt{2N} + {X_n^{(1)} \over 2^{1/2}N^{1/6}}; {t^{(1)} \over N^{1/3}};
\sqrt{2N} + {X_1^{(2)} \over 2^{1/2} N^{1/6}}, \dots,
\sqrt{2N} + {X_m^{(2)} \over 2^{1/2} N^{1/6}}; {t^{(2)} \over N^{1/3}} \Big )
\end{eqnarray}
Analogous to (\ref{n.2})--(\ref{n.5}), the matrix elements of
$\rho_{(n+m)}^{\rm soft}$ are computed from the formulas
\begin{eqnarray}\lefteqn{
S_1^{\rm soft}(X,Y;t_X,t_Y)} \nonumber \\ && =  \lim_{N \to \infty}
{e^{-N^{2/3}(t_Y - t_X)} \over 2^{1/2} N^{1/6}}
S \Big ( \sqrt{2N} + {X \over 2^{1/2} N^{1/6}},
 \sqrt{2N} + {Y \over 2^{1/2} N^{1/6}}; {t_X \over N^{1/3}},
 {t_Y \over N^{1/3}} \Big ) \label{r.2}
\end{eqnarray}
\begin{eqnarray}\lefteqn{ 
\tilde{S}_1^{\rm soft}(X,Y;t_X,t_Y)} \nonumber \\ &&  =  \lim_{N \to \infty}
{e^{-N^{2/3}(t_Y - t_X)} \over 2^{1/2} N^{1/6}}
\tilde{S} \Big ( \sqrt{2N} + {X \over 2^{1/2} N^{1/6}},
 \sqrt{2N} + {Y \over 2^{1/2} N^{1/6}}; {t_X \over N^{1/3}},
  {t_Y \over N^{1/3}} \Big ) \label{r.3}
\end{eqnarray}
\begin{eqnarray}\lefteqn{ 
\tilde{I}_1^{\rm soft}(X,Y;t_X,t_Y)} \nonumber \\ &&  =  \lim_{N \to \infty}
e^{N^{2/3}(t_Y + t_X)} \tilde{I} \Big ( \sqrt{2N} + {X \over 2^{1/2}
N^{1/6}},
 \sqrt{2N} + {Y \over 2^{1/2} N^{1/6}}; {t_X \over N^{1/3}},
   {t_Y \over N^{1/3}} \Big ) \label{r.4} 
\end{eqnarray}
\begin{eqnarray}\lefteqn{
D_1^{\rm soft}(X,Y;t_X,t_Y)} \nonumber \\ & & = \lim_{N \to \infty}  
{e^{-N^{2/3}(t_Y + t_X)} 
 \over 2 N^{1/3}}
D \Big ( \sqrt{2N} + {X \over 2^{1/2}
N^{1/6}},
 \sqrt{2N} + {Y \over 2^{1/2} N^{1/6}}; {t_X \over N^{1/3}},
{t_Y \over N^{1/3}} \Big ). \label{r.5}
\end{eqnarray}
Here the scale factors in addition to $1/2^{1/2} N^{1/6}$ do not effect
the value of (\ref{2.s}) because they result from a transformation of the
form (\ref{n.6'}).

\subsection{$\beta = 1$ initial conditions}

We see from (\ref{4.4}) that for $N$ even and $p \geq N$ 
\beqa
\beta_{p l} = \beta_{p N-1}\, \beta_{N-1 l} \,  .
\label{4.7}
\eeqa
Substituting in 
the formula (\ref{2.1e}) for $S_{2}$ and using (\ref{2.l}) and (\ref{2.m}) 
we thus have 
\begin{eqnarray}
S_2(x,y;\tau_x,\tau_y) & = & \sqrt{w(y)} \Big (
{\Phi_{N-2}(x;\tau_x) \over r_{N/2 - 1}(0)} e^{\gamma_{N-2} \tau_x} -
\sqrt{w(x)} {C_{N-1}(x) e^{-\gamma_{N-1} \tau_x} \over h_{N-1}} \Big )
\nonumber \\
&& \times C_{N-1}(y) e^{\gamma_{N-1} \tau_y}.
\label{4.8}
\eeqa

To compute the scaled limit we make use of the asymptotic expansion \cite{Sz}
\beqa
\label{4.9}
e^{-x^2/2} H_n(x) = \pi^{-3/4} 2^{n/2 + 1/4} (n!)^{1/2} n^{-1/12}
\Big ( \pi {\rm Ai}(-u) + {\rm O}(n^{-2/3}) \Big )
\eeqa
where $x = (2n)^{1/2} - u/2^{1/2}n^{1/6}$. This gives 
\beqa
\label{4.10}
\sqrt{w(y)} C_{N-1}(y) \Big |_{y = (2N)^{1/2} + Y/2^{1/2} N^{1/6}} 
\sim \pi^{1/4} 2^{-(N-1)/2 + 1/4}(N-1)!^{1/2} N^{-1/12}
{\rm Ai}(Y).
\eeqa
Also, from (\ref{2.m}), (\ref{4.1}) and (\ref{4.4}) 
\begin{eqnarray*}
f(x;\tau_x) & := & {\Phi_{N-2}(x;\tau_x) \over r_{N/2 - 1}(0)}
e^{\gamma_{N-2} \tau_x} - \sqrt{w(x)} {C_{N-1}(x) e^{-\gamma_{N-1} \tau_x}
\over h_{N-1}} \\
& = & {e^{-x^2/2} \over (N/2 - 1)!} e^{-N \tau_x}
\sum_{p=0}^\infty {(N/2 + p)! \over \sqrt{\pi} (N + 2p + 1)!}
H_{N + 2p + 1}(x) e^{-(2p + 3/2) \tau_x}.
\nonumber
\end{eqnarray*}
Noting that for large $N$ 
\beqa
\nonumber 
{(N/2 + p)! \over ((N + 2p + 1)!)^{1/2}}
{((N-1)!)^{1/2} \over (N/2 - 1)!} \sim 2^{-(p+1)}
\eeqa
and making use of (\ref{4.10}) with $n = N + 2p + 1$, 
$-u \sim X - 2p/N^{1/3}$ shows 
\begin{eqnarray}
\label{4.11}
\lefteqn{((N-1)!)^{1/2} f \Big ( (2N)^{1/2} + {X \over 2^{1/2} N^{1/6}};
{t_X \over N^{1/3}} \Big ) } \nonumber
\\ && \sim e^{-N^{2/3} t} 2^{(N-1)/2} 2^{1/4} \pi^{-1/4} N^{-1/12}
{N^{1/3} \over 2} \int_0^\infty {\rm Ai}(X - v) e^{-vt_X} \, dv.
\end{eqnarray}
Hence 
\begin{eqnarray} \lefteqn{
S_2 \Big (  (2N)^{1/2} + {X \over 2^{1/2} N^{1/6}},
 (2N)^{1/2} + {Y \over 2^{1/2} N^{1/6}}; {t_X \over N^{1/3}},
 {t_Y \over N^{1/3}} \Big )} \nonumber \\ && \sim
 e^{N^{2/3}(t_Y - t_X)} 2^{-1/2} N^{1/6} {\rm Ai}(Y)
 \int_0^\infty {\rm Ai}(X - v) e^{-vt_X} \, dv.
\label{4.12}
\end{eqnarray}

Now consider $S_{1}$ as 
defined by (\ref{2.1d}). Substituting (\ref{4.1}) we see that the 
asymptotics 
can be determined by writing the sum so that the sum index $p$ is replaced 
by $N-p$, and then using Stirling's formula and (\ref{4.10}). We thus find 
\begin{eqnarray} \lefteqn{
S_1 \Big (  (2N)^{1/2} + {X \over 2^{1/2} N^{1/6}},
(2N)^{1/2} + {Y \over 2^{1/2} N^{1/6}}; {t_X \over N^{1/3}},
{t_Y \over N^{1/3}} \Big )} \nonumber \\ && \sim
e^{N^{2/3}(t_Y - t_X)} 2^{1/2} N^{1/6} 
\int_0^\infty {\rm Ai}(X + v) {\rm Ai}(Y + v) e^{-v(t_Y - t_X)} \, dv.
\label{4.13}
\end{eqnarray}
Adding together (\ref{4.12}) and (\ref{4.13}) and substituting for $S$ in
(\ref{r.2}) we thus have
\begin{equation}\label{4.13'}
S_1^{\rm soft}(X,Y;t_X,t_Y) = 
\int_0^\infty {\rm Ai}(X + v) {\rm Ai}(Y + v) e^{-v(t_Y - t_X)} \, dv
 + {1 \over 2}
 {\rm Ai}(Y)
  \int_0^\infty {\rm Ai}(X - v) e^{-vt_X} \, dv.
\end{equation}

\subsubsection*{Asymptotics of $\tilde{I}(x, y; \tau_{x}, \tau_{y})$}
We know from (\ref{2.p}) that $\tilde{I}$ is defined in terms of 
$\{\Phi_{l}(\cdot ; \tau)\}$. 
The asymptotics of $\Phi_{2k-2}$ is calculated in an analogous fashion to the 
expansion (\ref{4.11}). Thus after noting 
\begin{eqnarray*}
\lefteqn{\Phi_{N+2k-2}(x;\tau) e^{\gamma_{N+2k-2} \tau} } \\ &&
=
{r_{N/2 + k - 1}(0) \over (N/2 + k -1)!} e^{-N \tau}
\sum_{m=k-1}^\infty {H_{2m+N + 1}(x) (N/2 + m)! \over
\pi^{1/2} (N + 2m + 1)!} e^{-(2m+3/2) \tau}
\end{eqnarray*}
we find 
\begin{eqnarray}
\Phi_{N+2k-2}\Big ( (2N)^{1/2} + {X \over 2^{1/2} N^{1/6}};
{t_X \over N^{1/3}} \Big ) e^{\gamma_{N+2k-2} t_X/N^{1/3}} \sim
{r_{N/2 + k -1}(0) \over ((N-1)!)^{1/2}}
{(N/2 - 1)! \over (N/2 + k - 1)!}
 \nonumber \\
\times  e^{-N^{2/3} t_X}
2^{(N-1)/2} 2^{1/4} \pi^{-1/4} N^{-1/12} {N^{1/3} \over 2}
\int_{2k/N^{1/3}}^\infty {\rm Ai}(X - v) e^{-v t_X} \, dv.
\label{4.14}
\end{eqnarray}

Consider now $\Phi _{2k-1}$. We note from 
(\ref{2.m}), (\ref{4.1}) and (\ref{4.4}) that 
\beqa
\Phi_{N+2k-1}(y;\tau) e^{\gamma_{N+2k-1} \tau} = - \sqrt{w(y)}
r_{N/2 + k -1}(0) {H_{N + 2k - 2}(y) \over \pi^{1/2}
(N + 2k - 2)!} e^{-N \tau} e^{- (2k - 3/2) \tau}
\nonumber
\eeqa
so use of (\ref{4.9}) gives 
\begin{eqnarray}
\label{4.15}
\Phi_{N+2k-1}\Big ( (2N)^{1/2} + {Y \over 2^{1/2} N^{1/6}};
{t_Y \over N^{1/3}} \Big ) e^{\gamma_{N+2k-1} \tau_Y/N^{1/3}} \sim
- r_{N/2 + k -1}(0) e^{-N^{2/3} t_Y} \nonumber \\
\times e^{-2k t_Y/N^{1/3}} 
\pi^{-1/4} 2^{(N + 2k - 2)/2 + 1/4} ((N + 2k - 2)!)^{1/2} N^{-1/12}
{\rm Ai} (Y - {2k \over N^{1/3}}).
\end{eqnarray}
Multiplying (\ref{4.14}) and (\ref{4.15}) together, substituting in (\ref{2.p}) and making 
use of the explicit value of $r_{m}(0)$ from (\ref{4.2}), we then find
\beqa
\label{4.16}
\tilde{I}_1^{\rm soft}(X,Y;t_X,t_Y) =
\int_0^\infty ds \, {\rm Ai}(Y - s) e^{-s t_Y}
\int_s^\infty {\rm Ai}(X - v) e^{-vt_x} dv - ( X \leftrightarrow Y )
\eeqa

\subsubsection*{Asymptotics of $D(x, y; \tau_{x}, \tau_{y})$}

Here the asymptotics of $\{R_{m}(\cdot ; \tau)\}$ are required.  
Now, from (\ref{2.i}) 
and (\ref{4.2}) we have 
\beqa
\nonumber 
R_{N - 2l - 2}(x;\tau) e^{\gamma_{N - 2l - 2} \tau} = 2^{-(N - 2l - 2)}
H_{N - 2l - 2}(x) e^{\gamma_{N - 2l - 2} \tau}.
\eeqa
Use of (\ref{4.9}) then gives
\begin{eqnarray}
\label{4.17}
\lefteqn{
\sqrt{w(x)} R_{N - 2l - 2}(x;\tau) e^{\gamma_{N - 2l - 2} \tau}
\bigg |_{\begin{array}{l} \scriptstyle x = (2N)^{1/2} + X/2^{1/2} N^{1/6}  \\ 
 \scriptstyle \tau = t_X / N^{1/3}\end{array}} \sim e^{N^{2/3} t_X}} \nonumber\\
&& \times  e^{-2l t_X /N^{1/3}} \pi^{1/4}
2^{-(N - 2l - 2)/2} 2^{1/4} ((N - 2l - 2)! )^{1/2} N^{-1/12}
{\rm Ai} \Big ( X + {2l \over N^{1/3}} \Big ).
\end{eqnarray}
Furthermore
\begin{eqnarray}
\label{4.17'}
R_{2m+1}(x;\tau) e^{\gamma_{2m+1} \tau} & = &
2^{-(2m+1)} H_{2m+1}(x) e^{\gamma_{2m+1} \tau} - m
2^{-(2m-1)} H_{2m - 1}(x) e^{\gamma_{2m - 1} \tau} \nonumber
\\ & = &
- e^{\gamma_{2m + 1} \tau} e^{x^2 / 2}
{d \over dx} \Big ( e^{-x^2 / 2} 2^{-2m} H_{2m}(x) \Big ) \nonumber \\
&&
- m 2^{-(2m - 1)} H_{2m - 1}(x) e^{\gamma_{2m - 1} \tau}
( 1 - e^{2 \tau} ),
\end{eqnarray}
and use of (\ref{4.9}) gives for the asymptotics 
\begin{eqnarray}
\label{4.18}
\lefteqn{
\sqrt{w(x)} R_{N - 2l - 1}(y;\tau) e^{\gamma_{N - 2l - 1} \tau}
\bigg |_{\begin{array}{l} \scriptstyle
y = (2N)^{1/2} + Y/2^{1/2} N^{1/6}  \\ 
 \scriptstyle
\tau = t_Y / N^{1/3}\end{array}} \sim e^{N^{2/3}t_Y}} \nonumber \\
&& \times \pi^{1/4} 2^{-(N - 2l - 2)/2} 
2^{3/4} ((N - 2l - 2)!)^{1/2} N^{1/12}
{d \over ds} \Big ( e^{-s t_Y} {\rm Ai} (Y + s) \Big ), 
\end{eqnarray}
$s:= 2l / N$.
Multiplying 
together (\ref{4.17}) and (\ref{4.18}), dividing by $r_{(N-2l-2)/2}$ and 
summing over $l$ (which is of the form of a Riemann integral) we thus obtain 
\beqa
\label{4.19}
D_1^{\rm soft}(X,Y;t_X,t_Y) = {1 \over 4} \int_0^\infty ds \,
e^{-st_X} {\rm Ai}(X + s) {d \over ds} \Big ( e^{-st_Y}
{\rm Ai}(Y + s) \Big ) - (X \leftrightarrow Y)
\eeqa

\subsubsection*{Asymptotics of $g(x, y; \tau_{y} - \tau_{x})$}

According to (\ref{2.3}) and (\ref{2.4}) 
\beqa
\label{4.19'}
g(x,y;\tau) = e^{-(x^2 + y^2)/2}
\sum_{j=0}^\infty {H_j(x) H_j(y) e^{ - (j + 1/2) \tau} \over 
\pi^{1/2} 2^j j!}
\eeqa
This summation can be evaluated in closed form \cite{Er} giving 
\beqa
\nonumber
g(x,y;\tau) = e^{-(x^2 + y^2)/2}
{e^{-\tau/2} \over (1 - e^{-2\tau})^{1/2}}
\exp \Big ( - {e^{-2 \tau} \over 1 - e^{-2 \tau} } (x - y)^2 \Big )
\exp \Big ( {2 e^{-\tau} \over 1 - e^{-2\tau} } xy \Big )
\eeqa
Hence
\beqa
\nonumber
g \Big ( (2N)^{1/2} + {X \over \sqrt{2} N^{1/6}},
(2N)^{1/2} + {Y \over \sqrt{2} N^{1/6}}; {t \over N^{1/3}} \Big )
\sim {N^{1/6} \over (2t)^{1/2}}
e^{- t N^{2/3}} e^{-(X-Y)^2/4t - t(X+Y)/2}.
\eeqa
Alternatively, proceeding as in the derivation 
of (\ref{4.13}), we can deduce from 
(\ref{4.19'}) that
\beqa
\label{4.19''}\lefteqn{
g \Big ( (2N)^{1/2} + {X \over \sqrt{2} N^{1/6}},
(2N)^{1/2} + {Y \over \sqrt{2} N^{1/6}}; {t \over N^{1/3}} \Big )}
\nonumber \\ &&
\sim e^{- N^{2/3} t} 2^{1/2} N^{1/6} \int_{-\infty}^\infty
{\rm Ai}(X + v) {\rm Ai}(Y + v) e^{vt} \, dv
\eeqa
Dividing by $ e^{- N^{2/3} t}
2^{1/2} N^{1/6}$, replacing $t$ by $t_X - t_Y$
 and subtracting from (\ref{4.13'}) shows 
\beqa
\tilde{S}_1^{\rm soft}(X,Y;t_X,t_Y) & = & -
 \int_{-\infty}^0 
 {\rm Ai}(X + v) {\rm Ai}(Y + v) e^{-v(t_Y - t_X) } \, dv \nonumber \\
&& +
 {1 \over 2} {\rm Ai}(Y) \int_0^\infty {\rm Ai} (X - v) e^{-vt_X} \, dv.
\eeqa

\subsection{$\beta = 4$ initial conditions}

Here we want to again compute the scaled distribution (\ref{r.1}). 
As for $\beta = 4$ 
initial 
conditions in the Laguerre ensemble, we must modify the 
scale factor in (\ref{r.4}) 
and (\ref{r.5}) so that 
\begin{eqnarray}
\label{a.1}\lefteqn{
\tilde{I}_4^{\rm soft}(X,Y;t_X,t_Y)} \nonumber \\
&&
 = \lim_{N \to \infty}
{e^{N^{2/3}(t_X + t_Y)} \over 2 N^{1/3}}
\tilde{I} \Big ( (2N)^{1/2} + {X \over \sqrt{2} N^{1/6}},
(2N)^{1/2} + {Y \over \sqrt{2} N^{1/6}}; {t \over N^{1/3}} \Big ) 
\end{eqnarray}
\begin{eqnarray}\lefteqn{
{D}_4^{\rm soft}(X,Y;t_X,t_Y)} \nonumber \\ &&  = \lim_{N \to \infty}
e^{-N^{2/3}(t_X + t_Y)} 
D \Big ( (2N)^{1/2} + {X \over \sqrt{2} N^{1/6}},
(2N)^{1/2} + {Y \over \sqrt{2} N^{1/6}}; {t \over N^{1/3}} \Big )
\label{a.2}
\eeqa

Substituting  
(\ref{4.5}) in (\ref{2.1e}) shows that in terms of $\{C_{l}(x)\}$, $S_{2}$ 
only consists of a single term, which reads explicitly 
\beqa
\label{4.2a}
S_2(x,y;\tau_x,\tau_y) = \sqrt{w(x) w(y)}
{C_N(x) e^{-\gamma_N \tau_x} \over h_N}
\Big ( - {N \over 2} \Big ) R_{N-2}(y;\tau_y) e^{\gamma_{N-2} \tau_y}.
\eeqa
The explicit form of $R_{N-2}(y; \tau_{y}) e^{\gamma_{N-2} \tau_{y}}$ 
is given by 
(\ref{4.3}). 
This expression however does not give a Riemann sum after substituting the 
asymptotic form of the integrand. Instead, we make use of the identity 
\beqa
\label{4.3a}
\sum_{l=0}^m {m! \over l!} 2^{-2l} H_{2l}(x) c^{-l} = - \alpha
(4c)^{-m} e^{\alpha x^2} \int_{-\infty}^x e^{-\alpha s^2}
H_{2m+1}(s) \, ds,
\eeqa
$c = (1 - \alpha)/\alpha$.  This can be verified by multiplying both sides by 
$e^{-\alpha x^{2}}$ and differentiating 
(we also verify that both sides have the same 
$x \rightarrow \infty$ behaviour).  
Choosing $c = e^{-2\tau}$ and comparing with 
(\ref{4.3}) we see that 
\begin{eqnarray}
R_{2m}(y;\tau) e^{\gamma_{2m} \tau} & = &
- {e^{\tau/2} \over 1 + e^{-2 \tau} } (2 e^{-\tau})^{-2m}
e^{y^2/ (1 + e^{-2 \tau})} \int_{-\infty}^y
e^{-s^2/ (1 + e^{-2 \tau})} H_{2m+1}(s) \, ds \nonumber \\
& = &
 {e^{\tau/2} \over 1 + e^{-2 \tau} } (2 e^{-\tau})^{-2m}
 e^{y^2/ (1 + e^{-2 \tau})} \int_{y}^\infty
 e^{-s^2/ (1 + e^{-2 \tau})} H_{2m+1}(s) \, ds
\label{4.3b}
\eeqa
where to obtain the second equality the fact 
that the integrand is odd has been used. 

With $x = (2N)^{1/2} + X/2^{1/2}N^{1/6}$, making the change of variables 
$s \mapsto (2N)^{1/2} + s$ 
and using the asymptotic expansion (\ref{4.9}) we see that 
\begin{eqnarray}
\sqrt{w(y)} R_{N - 2m}\Big ((2N)^{1/2} + {Y \over 2^{1/2} N^{1/6}};
{t \over N^{1/3}} \Big ) e^{\gamma_{N - 2m} t / N^{1/3}} \sim
2^{-3/4} N^{-1/4} \pi^{1/4} 2^{-(N - 2m)/2}  \nonumber
\\  
\times ((N - 2m + 1)!)^{1/2}
e^{N^{2/3} t} e^{Yt} \int_{Y+u}^\infty e^{-st} {\rm Ai}(s) \, ds,
\quad u := 2m/N^{1/3}
\label{4.3c}
\eeqa
For the $x$-dependent 
factor in (\ref{4.2a}), the asymptotic form of $\sqrt{w(x)} C_{N}(x)$ 
is given by 
(\ref{4.10}) with $N - 1$ replaced by $N$. Making use of the explicit value of 
$h_{N}$, we thus find 
\beqa
\label{4.3d}
\lefteqn{S_2 \Big ((2N)^{1/2} + {X \over 2^{1/2} N^{1/6}},
(2N)^{1/2} + {Y \over 2^{1/2} N^{1/6}};
{t_X \over N^{1/3}}, {t_Y \over N^{1/3}} \Big )} \nonumber \\ &&
 \sim
- {N^{1/6} \over 2^{1/2}} e^{-N^{2/3} (t_X - t_Y)}
{\rm Ai}(X) e^{Y t} \int_Y^\infty e^{-st} {\rm Ai}(s) \, ds.
\eeqa

The asymptotic form of the quantity 
$S_{1}$ as defined by (\ref{2.1d}) is given by 
(\ref{4.13}), since $S_{1}$ is the same for $\beta = 1$ and $\beta = 4$ initial 
conditions. Hence 
\beqa \label{ls4}
S_4^{\rm soft}(X,Y;t_X,t_Y) & = &
\int_0^\infty {\rm Ai}(X + v) {\rm Ai}(Y + v) e^{-v(t_Y - t_X)} \,
dv \\ &&
- {1 \over 2} {\rm Ai}(X) e^{Yt} \int_Y^\infty e^{-st} {\rm Ai}(s) \, ds
\eeqa
Also, subtracting (\ref{4.19''}) divided by 
$e^{-N^{2/3} t}2^{1/2}N^{1/6}$ (and with $t \mapsto t_X - t_Y$) gives 
\beqa
\tilde{S}_4^{\rm soft}(X,Y;t_X,t_Y) & = &
- \int_{-\infty}^0 {\rm Ai}(X + v) {\rm Ai}(Y + v) e^{-v(t_Y - t_X)} \,
dv \nonumber \\
&&- {1 \over 2} {\rm Ai}(X) e^{Yt} \int_Y^\infty e^{-st} {\rm Ai}(s) \, ds
\eeqa

\subsubsection*{Asymptotics of $\tilde{I}(x, y; \tau_{x}, \tau_{y})$}

First note from (\ref{2.m}), (\ref{4.1}), (\ref{4.3}) and (\ref{4.5}) that 
\begin{eqnarray*}\lefteqn{
\Phi_{2k-1}(x;\tau) e^{\gamma_{2k-1}\tau}} \\
 &  &
 = \sqrt{w(x)} e^{-\gamma_{2k-2} \tau} 2^{-2(k-1)} \Big (
{1 \over 2} H_{2k}(x) e^{-2\tau} - (2k-1) H_{2k-2}(x) \Big ) 
\\ &  & = -  e^{-\gamma_{2k-2} \tau} 2^{-2(k-1)} \bigg (
{d \over dx} \Big ( e^{-x^2/2} H_{2k-1}(x) \Big ) - {\sqrt{w(x)} \over 2}
H_{2k}(x) (e^{-2\tau} - 1) \Big )
\end{eqnarray*}
(c.f.\ (\ref{4.17'})). Use of (\ref{4.9}) gives for the asymptotics 
\beqa
\label{4.3e}
\lefteqn{ \Phi_{N - 2k - 1} \Big ( (2N)^{1/2} + {X \over 2^{1/2} N^{1/6}};
{t \over N^{1/3}} \Big ) e^{\gamma_{N-2k-1}t/N^{1/3}} \sim} \nonumber \\&&
- e^{-N^{2/3} t_X} \pi^{1/4} 2^{7/4} 2^{-(N-2k-1)/2} (N - 2k - 1)!^{1/2}
N^{1/12} {d \over ds} \Big ( e^{st_X} {\rm Ai}(X + s) \Big ),
\eeqa
$s : = 2 k/N$. Similarly 
\beqa
\nonumber
{1 \over r_{k-1}(0)} \Phi_{2k-2}(y;\tau) e^{\gamma_{2k-2}\tau} =
\sqrt{w(y)}  {H_{2k-1}(y) e^{-\gamma_{2k-1}\tau} \over \pi^{1/2}(2k-1)!}
\eeqa
and so 
\beqa
\label{4.3f}\lefteqn{
{1 \over r_{N/2 -k - 1}(0)} \Phi_{N-2k-2}\Big (
(2N)^{1/2} +  {Y \over 2^{1/2} N^{1/6}};
{t \over N^{1/3}} \Big ) e^{\gamma_{N-2k-2}t/N^{1/3}} \sim} \nonumber \\&&
e^{-N^{2/3} t_Y} e^{st_Y} \pi^{-1/4}
2^{(N-2k-1)/2 + 1/4} (N - 2k - 1)!^{-1/2} N^{-1/12}
{\rm Ai}(Y + s).
\eeqa
Multiplying together 
(\ref{4.3e}) and (\ref{4.3f}) and summing according to (\ref{2.p}) we 
see that
\beqa
\label{4.3g}
\tilde{I}_4^{\rm soft}(X,Y;t_X,t_Y) =
\int_0^\infty ds \, e^{st_Y} {\rm Ai}(Y + s)
{d \over ds} \Big ( e^{st_X} {\rm Ai}(X + s) 
\Big ) - (X \leftrightarrow Y )
\eeqa

\subsubsection*{Asymptotics of $D(x, y; \tau_{x}, \tau_{y})$}

Here we require the asymptotics of $R_{2m}(\cdot ; \tau)$ and 
 $R_{2m+1}(\cdot ; \tau)$. The former is given by (\ref{4.3c}). For the latter, we note from 
(\ref{4.3}) that 
\beqa
\nonumber
{1 \over r_m(0)} R_{2m+1}(x;\tau) e^{\gamma_{2m+1} \tau} =
{H_{2m+1}(y) \over 2 \sqrt{\pi} \Gamma(2m+2) } e^{\gamma_{2m+2}
\tau}
\eeqa
and thus according to (\ref{4.9}) 
\beqa
\label{4.4a}
\lefteqn{
{\sqrt{w(x)} \over r_{N/2 - m}(0)} R_{N - 2m+1} (x;\tau)
e^{\gamma_{N - 2m + 1} \tau} \bigg |_{\begin{array}{l}
\scriptstyle
x = (2N)^{1/2} + X/2^{1/2} N^{1/6} \\ 
\scriptstyle \tau = t/N^{1/3} \end{array}} }
\\ && \sim {e^{N \tau_x} \over 2 \sqrt{\pi} }
{e^{-u t_X} \over (\Gamma ( N - 2m + 2))^{1/2}} \pi^{1/4}
2^{(N - 2m + 1)/2 + 1/4} N^{-1/12} {\rm Ai}(X + u).
\eeqa
Multiplying (\ref{4.4a}) and (\ref{4.3c}) according to (\ref{2.q}) we see that 
\begin{equation}
\label{4.4b}
D(X,Y;t_X,t_Y) = {1 \over 4} e^{N^{2/3}(t_X + t_Y)}
\int_0^\infty e^{-u t_Y} {\rm Ai}(Y + u) e^{X t_X}
\int_{X + u}^\infty e^{-st} {\rm Ai}(s) \, ds - (X \leftrightarrow Y). 
\end{equation}
%%%%%%%%%%% page 33 ends here %%%%%%%%%%%%
\section{Discussion}
\setcounter{equation}{0}

\subsection{Connection with $\beta = 2$ initial conditions}

We have calculated exact formulas for the dynamical distribution
function (\ref{2.c}) in the case of $\beta = 1$ and $\beta = 4$ initial
conditions, with the final state corresponding to a $\beta = 2$
equilibrium.  If in these formulas we take the limit $t_{X},
t_{Y} \rightarrow \infty$ with $t_{X} - t_{Y}$ fixed,
then we would expect no memory of the initial state to be retained,
but rather the distribution functions corresponding to a $\beta = 2$
initial condition (i.e.\ the equilibrium state) to result.

Consider first the hard edge.  Inspection of (\ref{3.20'}), (\ref{m.2}),
(\ref{m.3}) and (\ref{m.4}) shows that in this limit
\beqa
\lefteqn{S_{1}^{\rm{hard}}(X, Y; t_{X}, t_{Y}) \sim
S_{4}^{\rm{hard}}(X, Y,;t_{X}, t_{Y})} \nonumber \\
& &  \sim
S_{2}^{\rm{hard}}(X, Y,;t_{X}, t_{Y})
:= \frac{1}{4}\int_{0}^{1}J_{a}\left(\sqrt{uX}\right)
J_{a}\left(\sqrt{uY}\right)\, e^{-u(t_{X} - t_{Y})}
\,  du  \,  ,
\label{5.1} \\
\lefteqn{\tilde{S}_{1}^{\rm{hard}}(X, Y; t_{X}, t_{Y}) \sim
\tilde{S}_{4}^{\rm{hard}}(X, Y,;t_{X}, t_{Y})} \nonumber \\
& &  \sim
\tilde{S}_{2}^{\rm{hard}}(X, Y,;t_{X}, t_{Y})
:= -\frac{1}{4}\int_{1}^{\infty}
J_{a}\left(\sqrt{uX}\right)
J_{a}\left(\sqrt{uY}\right)\, e^{-u(t_{X} - t_{Y})}
\,  du   \,  ,
\label{5.2}
\eeqa
while from (\ref{3.22}), (\ref{3.35}), (\ref{3.23}) and (\ref{3.38})
we see that
\beqa
& & e^{(t_{X} + t_{Y})}\, \tilde{I}_{1}^{\rm{hard}}(X, Y; t_{X}, t_{Y})
 \sim e^{(t_{X} + t_{Y})}\, \tilde{I}_{4}^{\rm{hard}}(X, Y; t_{X}, t_{Y})
\sim 0  \,  ,
\label{5.3} \\
& & e^{-(t_{X} + t_{Y})}\, {D}_{1}^{\rm{hard}}(X, Y; t_{X}, t_{Y})
 \sim e^{-(t_{X} + t_{Y})}\, {D}_{4}^{\rm{hard}}(X, Y; t_{X}, t_{Y})
\sim 0  \,   .
\label{5.4}
\eeqa
Substituting these values in (\ref{2.s}) and then using (\ref{2.1b}),
and the formula 
$$\rm{Pf}(ZX) = (\rm{det}(ZX))^{1/2},$$
we see by
interchanges of particular rows and columns that $ZX$ is equivalent
to a block matrix of the form
\beqa
\left(\begin{array}{cc}
 0_{n+m} & A  \\
-A & 0_{n+m}  \end{array} \right) \, ,
\nonumber 
\eeqa
where 
\beqa
A  =  \left[  \begin{array}{cc}
\left[S_{2}^{\rm{hard}}\left(
X_{j}^{(1)}, X_{k}^{(1)}; t_{1}, t_{1}\right)\right]_{n \times n}
 & \left[\tilde{S}_{2}^{\rm{hard}}\left(
X_{j}^{(1)}, X_{k}^{(2)}; t_{1}, t_{2}\right)\right]_{n \times m}
 \\
\left[\tilde{S}_{2}^{\rm{hard}}\left(
X_{j}^{(2)}, X_{k}^{(1)}; t_{2}, t_{1}\right)\right]_{m \times n}
& \left[S_{2}^{\rm{hard}}\left(
X_{j}^{(2)}, X_{k}^{(2)}; t_{2}, t_{2}\right)\right]_{m \times m}
\end{array} \right]  \,    .
\label{5.5}
\eeqa
Thus the limiting form of the dynamical distribution is given by
\beqa
\rho_{(n+m)}\left(x_{1}^{(1)}, \ldots , x_{n}^{(1)}; \tau_{1};
x_{1}^{(2)}, \ldots , x_{m}^{(2)}; \tau_{2} \right) = \rm{det} \,A \,  .
\label{5.6}
\eeqa

The analysis in the soft edge case is very similar. The formulas
(\ref{5.3}) and (\ref{5.4}) again apply, as do the formulas
(\ref{5.1}) and (\ref{5.2}) with $S_{2}^{\rm{hard}}$,
$\tilde{S}_{2}^{\rm{hard}}$  replaced by
\beqa
S_{2}^{\rm{soft}}(X, Y; t_{X}, t_{Y}) &:=&
\int_{0}^{\infty}{\rm Ai}(X+v){\rm Ai}(Y+v)e^{-v(t_{Y}-t_{X})}dv \,  ,
\label{5.7} \\
\tilde{S}_{2}^{\rm{soft}}(X, Y; t_{X}, t_{Y}) &:=&
-\int_{-\infty}^{0}{\rm Ai}(X+v){\rm Ai}(Y+v)e^{-v(t_{Y}-t_{X})}dv \,  .
\label{5.8}
\eeqa
We remark that in the case $ n = m =1$, the resulting formula for
$\rho_{(n+m)}$ implied by (\ref{5.6}) and (\ref{5.5}) in both the
hard and soft edge theories agrees with that computed by
M\^{a}cedo \cite{Ma1,Ma296}.

\subsection{Connection between the soft and hard edges}

For the hard edge distribution function of the Laguerre  ensemble at
$\beta = 2$ and $\tau = 0$, it is known that by taking the limit
$a \rightarrow \infty$ and introducing new scaled coordinates,
the distribution function of the soft edge of the Gaussian ensemble
results \cite{Fo93}.  Specifically, the new scaled variable $x$ is
related to the scaled variable $X$ of the hard edge by
\beqa
X = a^{2} - 2\, a\, (a/2)^{1/3} \, x
\label{5.9}
\eeqa
and the distribution functions are related by
\beqa
& & \lim_{a \rightarrow \infty} \left(-2a(a/2)^{1/3}\right)^{n}
\rho_{(n)}^{\rm hard}\left(a^{2}-2a(a/2)^{1/3}x_{1}, \ldots ,
a^{2}-2a(a/2)^{1/3}x_{n} \right)
\nonumber \\
& & \quad \quad \quad \quad \quad \quad \quad \quad \quad \quad \quad
 =  \rho_{(n)}^{\rm soft}\left(x_{1}, \ldots , x_{n}\right)\, .
\label{5.10}
\eeqa
As noted in \cite{Fo93,BFP98} this property should persist for general
$\beta$. Further, with appropriate scaling of $\tau$, the result would be
expected to generalize to the dynamical distributions.

Let us first check this latter point for the dynamical distribution with a
$\beta = 2$ initial condition (\ref{5.6}). For this purpose we require the
asymptotic expansion \cite{Wa}
\beqa
J_{a}(x) \sim \left(\frac{2}{x}\right)^{1/3}
{\rm Ai}\left(\frac{2^{1/3}(a-x)}{x^{1/3}}\right)
\label{5.11}
\eeqa
valid for $a$ and $x$ large such that the argument of the Airy function is
of order unity.  Changing variables $u = 1 - w(2/a)^{2/3}$ in the integrals
of (\ref{5.1}) and (\ref{5.2}) and introducing the scaled coordinates
(\ref{5.9}), application of (\ref{5.11}) shows
\beqa
J_{a}\left(\sqrt{ux}\right) \sim \left(\frac{2}{a}\right)^{1/3} {\rm Ai}
(x + w) \, .
\label{5.11'}
\eeqa
Thus if we also scale $t$ according to
\beqa
t_{X} \mapsto (a/2)^{2/3} \,  t_{x}
\label{5.12}
\eeqa
then we see that
\beqa
& & \left(-2a(a/2)^{1/3}\right) S_{2}^{\rm hard}\left(a^{2} - 2a(a/2)^{1/3}x,
a^{2} - 2a(a/2)^{1/3}y; (a/2)^{2/3}t_{x}, (a/2)^{2/3}t_{y} \right)
\nonumber \\
& & \quad \quad \quad \quad \quad \quad \quad \quad \quad
 \sim  e^{-(a/2)^{2/3}(t_{x} - t_{y})} S_{2}^{\rm soft}(x, y; t_{x}, t_{y})\, ,
\nonumber \\
& & \left(-2a(a/2)^{1/3}\right) 
\tilde{S}_{2}^{\rm hard}\left(a^{2} - 2a(a/2)^{1/
3}x,
a^{2} - 2a(a/2)^{1/3}y; (a/2)^{2/3}t_{x}, (a/2)^{2/3}t_{y} \right)
\nonumber \\
& & \quad \quad \quad \quad \quad \quad \quad \quad \quad
 \sim    
e^{-(a/2)^{2/3}(t_{x} - t_{y})} S_{2}^{\rm soft}(x, y; t_{x}, t_{y})\, .
\nonumber
\eeqa
Since, as seen from (\ref{2.s}), the factors $ e^{-(a/2)^{2/3}(t_{x} - t_{y})}$
do not affect the value of $\rho_{(n+m)}$, we thus have the dynamical
extension of (\ref{5.10}),
\beqa
\lim_{a \rightarrow \infty}\left(-2a(a/2)^{1/3}\right)^{n+m} 
\rho_{(n+m)}^{\rm hard}\Big(a^{2} - 2a(a/2)^{1/3}x_{1},  \ldots ,
a^{2} - 2a(a/2)^{1/3}x_{n}; (a/2)^{2/3}t_{x}; && \nonumber \\
\qquad \qquad  a^{2} - 2a(a/2)^{1/3}y_{1}, \ldots , 
a^{2} - 2a(a/2)^{1/3}y_{m}; (a/2)^{2/3}t_{y} \Big ) && \nonumber \\ 
\qquad \qquad  =  \rho_{(n+m)}^{\rm hard}(x_{1}, \ldots, x_{n}; t_{x};
y_{1}, \ldots , y_{m}; t_{y}) \, . && 
\label{5.13}
\eeqa

Consider next $\beta = 1$ initial conditions. Inspection of (\ref{3.20'}),
(\ref{m.2}), (\ref{3.22}) and (\ref{3.23}) shows that the analysis used above
for $\beta = 2$ initial conditions again suffices to establish the
connection  formula.  For example, consider the formula (\ref{3.22}) for
$\tilde{I}_{1}^{\rm hard}$.  For large $a$ we see that
\beqa
\lefteqn{\tilde{I}_{1}^{\rm hard}(X, Y; t_{X}, t_{Y})}
\nonumber \\
& \sim & \frac{a^{2}}{4}
\int_{1}^{\infty} s^{-1 + a/2} J_{a}\left(\sqrt{sY}\right)
e^{-st_{Y}} \int_{s}^{\infty} \frac{e^{-ut_{X}}}{u}
J_{a}\left(\sqrt{uX}\right) du - (X \leftrightarrow Y)  .
\label{5.14}
\eeqa
Use of the 
asymptotic expansion (\ref{5.11'}) ( after making the appropriate change
of variables in the integrals) shows
\beqa
& & \left(-2a(a/2)^{1/3}\right) 
\tilde{I}_{1}^{\rm hard}\left(a^{2} - 2a(a/2)^{1
/3}x,
a^{2} - 2a(a/2)^{1/3}y; (a/2)^{2/3}t_{x}, (a/2)^{2/3}t_{y} \right)
\nonumber \\
& & \quad \quad \quad \quad \quad \quad \quad \quad \quad
 \sim   e^{-(a/2)^{2/3}(t_{x} + t_{y})} 
\tilde{I}_{1}^{\rm soft}(x, y; t_{x}, t_
{y}) \,  .
\nonumber
\eeqa
Similar formulas hold connecting $S_{1}^{\rm hard}$, $\tilde{S}_{1}^{\rm hard}$
and
$D_{1}^{\rm hard}$ to $S_{1}^{\rm soft}$, $\tilde{S}_{1}^{\rm soft}$ and
$D_{1}^{\rm soft}$ 
respectively, thus showing the formula (\ref{5.13}) holds for
$\beta = 1$ initial conditions. Repeating the analysis for $\beta = 4$ initial
conditions leads to the same conclusion.

\subsection{Connection with the bulk}

It is known \cite{Fo93} 
that after appropriate scaling of the variables, the static
distributions at the 
hard and soft edges tend to the bulk distributions at large
distances from the edge of the system. The required scaling is
\beqa
X \mapsto \alpha + 2\pi \sqrt{\alpha}\, \rho x\,  , \quad \quad
X \mapsto  -\left(\alpha + \pi \rho x /\sqrt{\alpha} \right) \,  ,
\label{5.15}
\eeqa
with 
$\alpha \rightarrow \infty$, for the hard and soft edges respectively. Here we
will show that this feature persists for the dynamical distributions.

Consider 
first the hard edge with $\beta =2$ initial conditions.  Changing variables
$u \mapsto  u^{2}$ and making use of the asymptotic expansion
\beqa
J_{a}(x) \mathop{\sim}\limits_{x \to \infty}
\left(\frac{2}{\pi x}\right)^{1/2} \cos(x - \pi a/2 - \pi /4)
\label{5.16}
\eeqa
in the formulas (\ref{5.1}) and (\ref{5.2}), we see that
\beqa
\lefteqn{\lim_{\alpha \rightarrow \infty} 
2\pi \sqrt{\alpha}\, \rho \, S_{2}^{\rm
 hard}
\left(\alpha + 2\pi \sqrt{\alpha}\, 
\rho x, \alpha + 2\pi \sqrt{\alpha}\, \rho y
; t_{x}, t_{y}
\right)} \quad \quad \quad \quad \quad
\nonumber \\
& \sim & \rho \int_{0}^{1}\cos(\pi u \rho (x-y))\, e^{-u^{2}(t_{x}-t_{y})}\, du
\, ,
\label{5.17} \\
\lefteqn{\lim_{\alpha \rightarrow \infty} 2\pi \sqrt{\alpha}\, \rho
\, \tilde{S}_{2}^{\rm hard} 
\left(\alpha + 2\pi \sqrt{\alpha}\, \rho x, \alpha +
 2\pi
\sqrt{\alpha}\, \rho y; t_{x}, t_{y}\right) }
\quad \quad \quad \quad \quad \nonumber \\
& = & -\rho \int_{1}^{\infty}\cos(\pi u \rho (x-y))\, e^{-u^{2}(t_{x}-t_{y})}\,
du \, .
\label{5.18}
\eeqa
Substituting these 
formulas in (\ref{5.5}) and (\ref{5.6}) and also making the 
change of
scale $t \mapsto (\pi \rho)^{2}t/2$ reclaims the known formula for the bulk
dynamical distributions \cite{NF98b}.

In fact the asymptotic behaviour 
(\ref{5.17}) and (\ref{5.18}) of $S_{\beta}^{\rm
 hard}$
and $\tilde{S}_{\beta}^{\rm hard}$ for $\beta = 2$ 
initial conditions persists for
$\beta
= 1$ and $\beta = 4$ 
initial conditions. This follows from application of (\ref{5.26})
in the exact results 
(\ref{3.20'}), (\ref{m.2}) and (\ref{m.3}), (\ref{m.4}). It
 remains
to consider 
$\tilde{I}_{\beta}$ and $D_{\beta}$. For $\beta = 1$ initial conditions
application of (\ref{5.16}) in (\ref{3.22}) and (\ref{3.23}) shows
\beqa
\lefteqn{\lim_{\alpha \rightarrow \infty} 2\pi \rho \, \tilde{I}_{1}^{\rm hard}
\left(\alpha + 2\pi \sqrt{\alpha}\, \rho x, \alpha + 2\pi \sqrt{\alpha}\,
\rho y; t_{x}, t_{y}\right)} \quad \quad \quad \quad \quad
\nonumber \\
& = & \rho \int_{1}^{\infty}du \,
\frac{e^{-u^{2}(t_{x}+t_{y})}}{u}\sin(\pi u \rho (y-x)) \, ,
\label{5.19} \\
\lefteqn{\lim_{\alpha \rightarrow \infty} 2\pi \rho \alpha \, D_{1}^{\rm hard}
\left(\alpha + 2\pi \sqrt{\alpha}\, 
\rho x, \alpha + 2\pi \sqrt{\alpha}\, \rho y
;
t_{x}, t_{y}\right)} \quad \quad \quad \quad \quad
\nonumber \\
& = & \rho \int_{0}^{1}du \,
u\, e^{u^{2}(t_{x}+t_{y})}\sin(\pi u \rho (y-x)) \, ,
\label{5.20}
\eeqa
while for $\beta = 4$ 
initial conditions application of (\ref{5.16}) in (\ref{3.35}) and
(\ref{3.38}) shows
\beqa
\lefteqn{\lim_{\alpha 
\rightarrow \infty} 2\pi \rho \alpha \, \tilde{I}_{4}^{\rm
 hard}
\left(\alpha + 2\pi 
\sqrt{\alpha}\, \rho x, \alpha + 2\pi \sqrt{\alpha}\, \rho y
;
t_{x}, t_{y}\right) } \quad \quad \quad \quad \quad
\nonumber \\
& = & \rho \int_{1}^{\infty}du \,
u\, e^{-u^{2}(t_{x}+t_{y})}\sin(\pi u \rho (y-x))  \, ,
\label{5.21} \\
\lefteqn{\lim_{\alpha \rightarrow \infty} 2\pi \rho \, D_{4}^{\rm hard}
\left(\alpha + 2\pi 
\sqrt{\alpha}\, \rho x, \alpha + 2\pi \sqrt{\alpha}\, \rho y
;
t_{x}, t_{y}\right) } \quad \quad \quad \quad \quad
\nonumber \\
& = & \rho \int_{0}^{1}du \,
\frac{e^{u^{2}(t_{x}+t_{y})}}{u}\sin(\pi u \rho (y-x))  \, .
\label{5.22}
\eeqa
After scaling $t \mapsto (\pi \rho)^{2}t/2$, the known formulas for
$\rho_{(1+1)}(x, y; t_{x}, t_{y})$ with 
$\beta = 1$ and $\beta = 4$ initial conditions
\cite{Fo96,FN97} are reproduced by these results.

In the case of the soft edge connecting to the bulk, similar formulas are found
to
hold. These formulas are derived from the asymptotic expansion
\beqa
{\rm Ai}(-x) \mathop{\sim}\limits_{x \to \infty}
\frac{1}{\pi^{1/2}x^{1/4}} \cos(2x^{3/2}/3 - \pi/4) \,  .
\label{5.23}
\eeqa
Consider for example $S_{2}^{\rm soft}$ as specified by (\ref{5.7}).  We first
change variables 
$v \mapsto \alpha(1-u)$ for $0 < v < \alpha$, and replace the Airy
functions in the integrand according to (\ref{5.23}). Noting that
\beqa
\frac{2}{3}\left\{
(\alpha + \pi \rho x /\sqrt{\alpha} - v)^{3/2} -
(\alpha + \pi 
\rho y /\sqrt{\alpha} - v)^{3/2}\right\} \sim \pi \rho u^{1/2}(x-y)\, ,
\nonumber
\eeqa
we see that if we also scale $t$ according to
\beqa
t \mapsto t/\alpha
\label{5.24}
\eeqa
then $S_{2}^{\rm soft}$ has the asymptotic form
\beqa
\lefteqn{\lim_{\alpha \rightarrow \infty} (-\pi \rho /\sqrt{\alpha})\,  
S_{2}^{\rm soft}
(-(\alpha + \pi \rho x /\sqrt{\alpha}), -(\alpha + \pi \rho y/\sqrt{\alpha});
t_{x}/\alpha , t_{y}/\alpha )} \quad \quad \quad \quad
\nonumber \\
& \sim & e^{-(t_{x}-t_{y})}
\rho \int_{0}^{1}\cos(\pi u \rho (x-y))\, e^{-u^{2}(t_{x}-t_{y})}\, du \, .
\label{5.25}
\eeqa
Apart from the factor of $e^{-(t_{x} - t_{y})}$ this is the same result as (\ref
{5.17}).
A similar calculation 
shows the asymptotic form of $\tilde{S}_{2}^{\rm soft}$ is
given by (\ref{5.18}) except for an additional factor of $e^{-(t_{x} -t_{y})}$.
This shows
that this additional factor cancels from the formula (\ref{2.s}) for the
distribution functions. 
The final result is that the bulk dynamical distributions
 are
reclaimed from the soft edge results for each of the three distinct initial
conditions. (As in the hard edge case a final rescaling $t \mapsto (\pi \rho)^{2
}t/2$
is required).

\subsection{Asymptotic  form of $\rho_{(1+1)}^{T}(X, Y; t)$}

The dynamical density-density correlation function between an eigenvalue  with
value $X$ initially, and an eigenvalue with value $Y$ and parameter $\tau$ is
defined in terms of the distribution function (\ref{2.c}) by
\beqa
\rho_{(1+1)}^{T}(x, y; \tau) = \rho_{(1+1)}(x; 0; y; \tau)
- \rho_{(1)}(x; 0) \, \rho_{(1)}(y; \tau) \,   .
\label{5.26}
\eeqa
According to (\ref{2.s}) and (\ref{2.1b}) we have that in the finite system
\beqa
\rho_{(1+1)}^{T}(x, y; \tau) = - \left( S(x, y; 0, \tau) \,
\tilde{S}(y, x; \tau, 0) - \tilde{I}(x, y; 0, \tau) \, D(x, y; 0, \tau) \right)
\,  .
\label{5.27}
\eeqa
For the Laguerre ensemble at the hard edge we define the corresponding scaled
correlation $\rho_{(1+1)}^{T \, \rm{hard}}$ by
\beqa\lefteqn{
\rho_{(1+1)}^{T \, \rm{hard}}(X, Y; t) =
\lim_{N \rightarrow \infty} \left( \frac{1}{4N}\right)^{2}  
\rho_{(1+1)}^{T}\left( \frac{X}{4N},  \frac{Y}{4N}; \frac{t}{2N} \right) }
\nonumber \\
&  & = - \left( S^{\rm hard}_{\beta}(X, Y; 0, t) \,
\tilde{S}_{\beta}^{\rm hard}(Y, X; t, 0)  
 - \tilde{I}_{\beta}^{\rm hard}(X, Y; 0, t) \,
D_{\beta}^{\rm hard}(X, Y; 0, t) \right) \,  ,
\label{5.28}
\eeqa
where $\beta = 1, 2$ or 4 depending on the initial conditions, while for the
Gaussian ensemble at the soft edge the scaled correlation
$\rho_{(1+1)}^{T \, \rm{soft}}$ is defined by
\beqa
\rho_{(1+1)}^{T \, \rm{soft}}(X, Y; t) & = &
\lim_{N \rightarrow \infty} \left( \frac{1}{2^{1/2}N^{1/6}}\right)^{2}
\rho_{(1+1)}^{T}\left( \sqrt{2N} + \frac{X}{2^{1/2}N^{1/6}},
\sqrt{2N} + \frac{Y}{2^{1/2}N^{1/6}}; \frac{t}{N^{1/3}} \right)
\nonumber \\
& = & - \left( S^{\rm soft}_{\beta}(X, Y; 0, t) \,
\tilde{S}_{\beta}^{\rm soft}(Y, X; t, 0) -
\tilde{I}_{\beta}^{\rm soft}(X, Y; 0, t) \,
D_{\beta}^{\rm soft}(X, Y; 0, t) \right)   .
\nonumber \\
\label{5.29}
\eeqa
The asymptotic 
form of $\rho_{(1+1)}^{\rm{hard}}(X, Y; t)$ for large $|X - Y|$ and $t$
is of interest in applications of the parameter dependent Laguerre ensemble to
transport in mesoscopic systems \cite{Ma296}.  To deduce this asymptotic form, a
hydrodynamical approximation to the corresponding Fokker-Planck equation was
made which implied
\beqa
\int_{0}^{\infty} \int_{0}^{\infty} (4XY) \,
\rho_{(1+1)}^{T \, {\rm hard}}\left( X^{2}, Y^{2}; t \right) \,
\cos(kX) \, \cos(kY) \, dX \, dY
\mathop{\sim} \limits_{k \to 0 \atop t \to \infty}
\frac{|k|}{2\beta}\, e^{-2t|k|}
\label{5.30}
\eeqa
(eq.\ (80) of \cite{Ma296} with $\pi \rho \delta u^{2}/\gamma = 2t$).
On the other hand we can easily check that
\beqa
{\rm Re}\left( \frac{1}{(v + u + 2it)^{2}} + \frac{1}{(v - u - 2it)^{2}} \right)
= -2 \int_{0}^{\infty} dk \, ke^{-2kt} \cos(kv) \cos(ku) \,  .
\label{5.30'}
\eeqa
Taking the inverse transform of this result and comparing with (\ref{5.30})
shows that the latter formula is equivalent to
\beqa
(4XY) \rho_{(1+1)}^{T \, {\rm hard}} \left(X^{2}, Y^{2}; t \right)
\mathop{\sim}\limits_{X,Y,t \to \infty}
-\frac{1}{\pi^{2} \beta}
{\rm Re}\left( \frac{1}{(X + Y + 2it)^{2}} + \frac{1}{(X - Y + 2it)^{2}} \right)
\label{5.31}
\eeqa
for the leading non-oscillatory behaviour.

The prediction 
(\ref{5.1}) can readily be checked on the exact formula (\ref{5.28})
with $\beta = 1, 2$ and 4 initial conditions.  The simplest case is $\beta = 2$
initial conditions, when the product
$\tilde{I}_{\beta}^{\rm hard}\tilde{D}_{\beta}^{\rm hard}$ 
in (\ref{5.28}) vanishes.
To analyze $S_{2}^{\rm hard}$  and
$\tilde{S}_{2}^{\rm hard}$ 
we make use of the asymptotic expansion (\ref{5.16}),
and furthermore expand the integrands about their endpoint $u = 1$. Integration
by
parts then gives
\beqa
S_{2}^{\rm hard}\left(X^{2}, Y^{2}; 0, t \right)
& \sim & \frac{e^{t}}{2\pi (XY)^{1/2}} \, {\rm Re}
\left(\frac{e^{i(X-Y)}}{-i(X-Y) - 2t} +
\frac{e^{i((X+Y) - \pi \alpha /2 - \pi/2)}}{-i(X+Y) - 2t} \right) \, ,
\nonumber \\
\label{5.32} \\
\tilde{S}_{2}^{\rm hard}\left(Y^{2}, X^{2}; t, 0 \right)
& \sim & \frac{e^{-t}}{2\pi (XY)^{1/2}} \, {\rm Re}
\left(\frac{e^{i(X-Y)}}{i(X-Y) - 2t} +
\frac{e^{i((X+Y) - \pi \alpha /2 - \pi/2)}}{i(X+Y) - 2t} \right) \, ,
\nonumber \\
\label{5.33}
\eeqa
and these formulas when substituted in (\ref{5.28}) imply the result (\ref{5.31}
)
with $\beta = 2$. 
In the case of $\beta = 1$ and $\beta = 4$ initial conditions,
similar analysis shows
\beqa
\lefteqn{\tilde{I}_{\beta}^{\rm hard}\left(X^{2}, Y^{2}; 0, t \right)
D_{\beta}^{\rm hard}\left(X^{2}, Y^{2}; 0, t \right)} \quad \quad \quad \quad
\nonumber \\
& \sim & \chi_{\beta} \tilde{S}_{\beta}^{\rm hard}\left(X, Y; 0, t \right)
\tilde{S}_{\beta}^{\rm hard}\left(Y, X; t, 0 \right)
\nonumber \\
& \sim &  \chi_{\beta}\frac{1}{4XY}\, \frac{1}{2\pi^{2}}
{\rm Re}
\left( \frac{1}{(X + Y + 2it)^{2}} + \frac{1}{(X - Y + 2it)^{2}} \right),
\label{5.34}
\eeqa
where 
$\chi_{\beta} = -1$ for $\beta = 1$ and 
$\chi_{\beta} = 1/2$ for $\beta = 4$,  and
the asymptotics 
refer to the leading non-oscillatory term. Thus (\ref{5.31}) again
remains valid.

Let us now turn our 
attention to the same asymptotic limit for the soft edge. There
it is known that the asymptotic form (\ref{5.31}) for $X, Y \rightarrow \infty$
is
valid in 
the static theory ($t = 0$) \cite{JF94}. It turns out that the exact results
for (\ref{5.29}) satisfy (\ref{5.31}) for $X, Y \rightarrow \infty$ and $t
\rightarrow \infty$, as just demonstrated for the hard edge correlations. Thus,
for
$\beta = 2$ initial conditions, 
use of the asymptotic expansion  (\ref{5.23}) in
(\ref{5.7}) and (\ref{5.8}) shows
\beqa
\lefteqn{S_{2}^{\rm soft}\left(-X^{2}, -Y^{2}; 0, t \right) } \quad \quad \quad
\nonumber \\
& \sim & \frac{1}{2 \pi (XY)^{1/2}} \, {\rm Re}
\left(\frac{e^{2i(X^{3}+Y^{3})/3 - iv(X+Y) - \pi i/2}}{i(X+Y) + t} +
\frac{e^{2i(X^{3}-Y^{3})/3 - iv(X-Y)}}{i(X-Y) + t}\right) \, ,
\nonumber \\
\lefteqn{\tilde{S}_{2}^{\rm soft}\left(-Y^{2}, -X^{2}; t, 0 \right)} 
\quad \quad
\quad
\nonumber \\
& \sim & \frac{1}{2 \pi (XY)^{1/2}} \, {\rm Re}
\left(\frac{e^{2i(X^{3}+Y^{3})/3 - iv(X+Y) - \pi  i/2}}{i(X+Y) - t} +
\frac{e^{2i(X^{3}-Y^{3})/3 - iv(X-Y)}}{i(X-Y) - t}\right) \, ,
\nonumber
\eeqa
and these formulas when substituted in (\ref{5.29}) (with
$\tilde{I}_{2}^{\rm soft}  D_{2}^{\rm soft} = 0$) gives the asymptotic form
(\ref{5.31}) (although $t$ therein must be rescaled by replacing $2t$ with $t$).
The same conclusion is reached for $\beta = 1$ and 4 initial conditions, where
we find the analogue of the formula (\ref{5.34}).

As pointed out 
in \cite{Ma296}, a consequence of the asymptotic form (\ref{5.31})
is a universal formula 
for the time displaced covariance of two linear statistics
\beqa
A_{t} = \sum_{j=1}^{N} a\left( x_{j}^{2}(t)\right) \, , \quad \quad
B_{t} = \sum_{j=1}^{N} b\left( x_{j}^{2}(t)\right) \,  .
\nonumber
\eeqa
Thus, if in the general formula
\beqa
{\rm Cov}(A_{0}, B_{t}) = \int_{0}^{\infty}dx \, a(x^{2}) \,
\int_{0}^{\infty}dy \, b(y^{2}) (4xy) \rho_{(1+1)}^{T}(x^{2}, y^{2}; t)
\nonumber
\eeqa
we suppose $a(x^{2})$ and $b(x^{2})$ are made slowly varying by writing
\beqa
a(x^{2}) \mapsto a((x/\alpha)^{2}) \, , \quad \quad
b(x^{2}) \mapsto b((x/\alpha)^{2}) \, ,
\nonumber
\eeqa
where $\alpha \gg 1$, and if we scale $t$ according to
\beqa
t \mapsto \alpha \, t \,   ,
\nonumber
\eeqa
then a simple change of variables and use of (\ref{5.31}) shows that for
$\alpha \rightarrow \infty$
\beqa
{\rm Cov}(A_{0}, B_{t}) & \sim & -\frac{1}{\pi^{2}\beta}
\int_{0}^{\infty} dx \, a(x^{2}) \int_{0}^{\infty} dy \, b(y^{2})
{\rm Re} \left( \frac{1}{(X+Y+2it)^{2}} + \frac{1}{(X-Y+2it)^{2}}\right)
\nonumber \\
& = & \frac{2}{\pi^{2} \beta} \int_{0}^{\infty} dk \, k \,
\tilde{a} (k) \,  \tilde{b} (k) \,  e^{-2kt} \,  ,
\label{5.35}
\eeqa
where the equality follows from (\ref{5.30'}), and
\beqa
\tilde{a} (k) := \int_{0}^{\infty} a(x^{2}) \cos(kx)\, dx
\nonumber
\eeqa
and similarly the meaning of $\tilde{b} (k)$.

\subsection{Empirical density of eigenvalues at
the soft edge}
The dynamical distribution functions at the soft edge calculated in 
Section 4 can be realized by parameter dependent random matrices
of the type (\ref{1.3}). Thus we can numerically construct random
matrices and empirically compute eigenvalue distributions for comparison
against the theoretical prediction. The simplest distribution to
compute empirically is the density (one-point function). Here we
will undertake such a calculation for the orthogonal to unitary
transition at the soft edge.

As deduced from (\ref{2.s}), (\ref{2.1b}) and (\ref{3.20'})
the theoretical prediction for the scaled density at the
soft edge of the  orthogonal to unitary transition is
\begin{eqnarray}\label{6.1}
\rho_{(1)}^{\rm soft}(X;t) & = & S_1^{\rm soft}(X,X;t,t) \nonumber \\
& = & \int_0^\infty \Big ( {\rm Ai} (X + v) \Big )^2 \, dv +
{1 \over 2} \int_0^\infty e^{-st_X} {\rm Ai} (X - s) \, ds.
\end{eqnarray}
The theory of Section 1 tells us that the parameter dependent random
matrices (\ref{1.3}), for an appropriate $X^{(0)}$ have, in
the infinite dimension limit, a
scaled density at the soft edge given by (\ref{6.1}).
Since we are considering the orthogonal to unitary
transition, $X^{(0)}$ must be chosen as
a real symmetric matrix with joint p.d.f.~for the elements
$$
P(X^{(0)}) = \prod_{j=1}^N \Big ( {1 \over 2 \pi} \Big )^{1/2}
e^{ - X_{jj}^{(0)\, 2}/2} \prod_{j < k} {1 \over \pi^{1/2}}
e^{ - X_{jk}^{(0) \, 2}}.
$$
The joint distribution of the elements of $X$, $P(X;\tau)$ say, is obtained
by integrating over $X_{jj}^{(0)}$ and $X_{jk}^{(0)}$. Thus
\begin{eqnarray}\label{6.7}
P(X;\tau) & = & \prod_{j=1}^N \int_{-\infty}^\infty dX_{jj}^{(0)} \,
\prod_{j < k}  \int_{-\infty}^\infty dX_{jk}^{(0)} \,
P(X^{(0)}) P(X^{(0)};X;\tau) \nonumber \\
& = &  \prod_{j=1}^N {e^{-X_{jj}^2}/(1 + e^{-2\tau}) \over
\sqrt{\pi (1 + e^{-2 \tau})}}
\prod_{j < k} {2 \over \pi \sqrt{1 - e^{- 4 \tau}}}
e^{- 2 ({\rm Re} X_{jk})^2/(1 + e^{-2\tau})}
e^{- 2 ({\rm Im} X_{jk})^2/(1 - e^{-2\tau})}, \nonumber \\
\end{eqnarray}
which tells us that the diagonal and the real and imaginary parts of the
upper triangular elements are all independent, having Gaussian distributions
with mean zero and variances $2/(1 + e^{-2 \tau})$, $4/(1 + e^{-2 \tau})$
and $4/(1 - e^{-2 \tau})$ respectively. Hence, for a given value of
$N$ and $\tau$ such matrices are simple to generate numerically. For
each such random matrix the eigenvalues $\lambda_j$ are computed and
scaled according to $\lambda \mapsto (\lambda_j - (2N)^{1/2})
2^{1/2} N^{1/6}$. For $N$ large enough and with $\tau$ related to $t$
by (\ref{4.6}) the corresponding empirical density should approach 
(\ref{6.1}). The situation for a relative small value of $N$ (recall
the occurence of $N^{1/6}$) is illustrated in Figure 1, where good
agreement with the infinite dimension limit is observed.

Also of interest is the possibility of providing a realization of the hard
edge results. Unfortunately here it seems that the initial conditions
used do not precisely correspond to a situation in which $X$ in
(\ref{1.6}) can be generated from a formula analogous to (\ref{6.7}).
The problem is that if $X^{(0)}$ is a real $n \times m$ matrix
with independent Gaussian entries,
then the
square of the eigenvalues of $X^{(0)\dagger} X^{(0)}$ has p.d.f.~(\ref{1.2}) 
with
$a = n - m - 1$. However in deducing (\ref{1.4}) from (\ref{1.6}) in
the case of transitions to unitary symmetry ($\beta = 2$) we must take
$a = n - m$, which is thus incompatible with the sought initial
condition.

\vspace{.5cm}
\begin{figure}
\epsfxsize=10cm
\centerline{\epsfbox{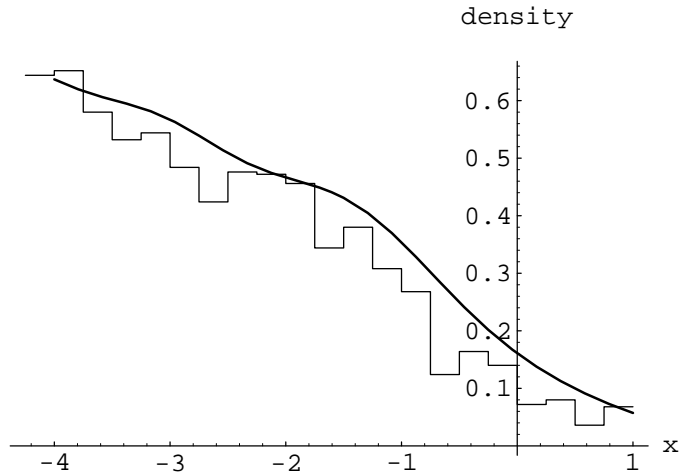}}
\caption{\label{1.f1} Empirical histogram of the scaled soft edge
density for  500 parameter dependent random matrices of the form
(\ref{6.7}) with $\tau = .05$ and $N = 100$, compared against
the infinite dimension form (\ref{6.1}). }
\end{figure}

\subsection{Distribution functions in the initial state}
With $m=0$ and $\tau_1 \to 0$ the dynamical distribution (\ref{1.7'})
reduces to the $n$-point distribution function in the initial state.
At $\beta = 1$ the formulas (\ref{2.s}), (\ref{2.1f}), (\ref{2.1g})
and (\ref{2.1h}) reclaim the well known \cite{MM91,FP95,NF95,TW98}
general formula
\begin{equation}\label{5.in}
\rho_{(n)}(x_1,\dots,x_n) =
{\rm Tdet} \left [ \begin{array}{cc}
S_1(x_j,x_k) & {1 \over 2} \int {\rm sgn}(x_k - y) S_1(x_j,y) \, dy -
{1 \over 2} {\rm sgn}(x_k - x_j) \\[.2cm]
{\partial \over \partial x_j} S_1(x_j,x_k) &
S_1(x_k,x_j) \end{array} \right ],
\end{equation}
where $S_1(x_j,x_k) := S(x_j,x_k;0,0)$ as specified by (\ref{2.o}) with
$\{R_j\}$ corresponding to the $\beta = 1$ initial condition.
We remark that the integral in (\ref{5.in}) can be rewritten
according to
$$
 {1 \over 2} \int {\rm sgn}(x_k - y) S_1(x_j,y) \, dy =
- \int_{x_j}^{x_k} S_1(x_j,y) \, dy.
$$
At $\beta = 4$ these same formulas (with (\ref{2.1h}) replaced by (\ref{2.1i}))
give that
\begin{equation}\label{5.in1}
\rho_{(n)}(x_1,\dots,x_n) =
{\rm Tdet} \left [ \begin{array}{cc}
S_4(x_j,x_k) & 
{\partial \over \partial x_k} S_1(x_k,x_j) \\[.2cm]
- \int_{x_k}^{x_j} S_4(y,x_k) \, dy &
S_4(x_k,x_j) \end{array} \right ],
\end{equation}
where $S_4(x,y) := S(x,y;0,0)$ as specified by (\ref{2.o}) with $\{R_j\}$
therein corresponding to the $\beta = 4$ initial condition, 
again  in accordance with the known result.
In deriving
(\ref{5.in1}) from (\ref{2.s}) we assumed that none of the points
$x_1,\dots,x_n$ are coincident; this is necessary because of the delta
functions in (\ref{2.b}) with $\beta = 4$.

We can take the limit $t_X,t_Y \to 0$ in the formulas obtained for
$S_1^{\rm edge}(X,Y;t_X,t_Y)$ (here edge denotes hard or soft) obtained
in Section 3 and 4 to obtain the explicit form of $S_1^{\rm edge}(X,Y)$
in the scaled limit at the hard and soft edges.
Thus from (\ref{3.20'}), (\ref{m.3}), (\ref{4.13'}) and (\ref{ls4})
we find
\begin{eqnarray}
S_1^{\rm hard}(X,Y) & =  & K^{\rm hard}(X,Y) -
{J_{a+1}(\sqrt{Y}) \over 4 \sqrt{Y}} \Big (
\int_0^{\sqrt{X}} J_{a+1}(u) \, du - 1 \Big ) \label{te1}\\
S_4^{\rm hard}(X,Y) & =  & K^{\rm hard}(X,Y) -
{1 \over 4 X^{1/2}} J_{a-1}(X^{1/2}) \int_0^{Y^{1/2}} J_{a+1}(s) \, ds
\label{te2} \\
S_1^{\rm soft}(X,Y)  & =  & K^{\rm soft}(X,Y)
+ {1 \over 2} {\rm Ai}(Y) \Big ( 1 - \int_X^\infty {\rm Ai}(t) \, dt \Big )
\label{te3} \\
S_4^{\rm soft}(X,Y)  & =  & K^{\rm soft}(X,Y) -
{1 \over 2} {\rm Ai}(X) \int_Y^\infty {\rm Ai}(s) \, ds \label{te4}
\end{eqnarray}
where
\begin{eqnarray}
K^{\rm hard}(X,Y) & := & {X^{1/2} J_{a+1}(X^{1/2}) J_a(Y^{1/2}) -
Y^{1/2} J_{a+1}(Y^{1/2}) J_a(X^{1/2}) \over 2 (X - Y) } \nonumber \\
 K^{\rm soft}(X,Y)  & := & { {\rm Ai}(X) {\rm Ai}'(Y) -
{\rm Ai}(Y) {\rm Ai}'(X) \over X - Y}.
\end{eqnarray}
In obtaining these formulas we have used (\ref{5.2}) and (\ref{5.7})
together with the known facts that \cite{Fo93}
$$
S_2^{\rm hard}(X,Y) = K^{\rm hard}(X,Y), \quad
S_2^{\rm soft}(X,Y) =  K^{\rm soft}(X,Y)
$$
to rewrite the first of the integrals in the expressions
 (\ref{3.20'}), (\ref{m.3}), (\ref{4.13'}) and (\ref{ls4}).
Also, Bessel function identities have been used in obtaining the second
term in (\ref{te1}) and (\ref{te2}), while in the second term of
(\ref{te3}) use has been made of the definite integral
$$
\int_{-\infty}^\infty {\rm Ai}(x) \, dx = 1.
$$
We remark that in previous studies \cite{NF95, Ve94} expression have
been obtained for (\ref{te1}) and (\ref{te2}) which are of a
more complicated structure. Also the expressions obtained in
\cite{Fo93} for (\ref{te3}) and (\ref{te4}) are now seen to be
in error.

Some caution must be exercised in applying these results at $\beta = 4$.
In particular, the one body weight function used in (\ref{2.b}) at
$\beta = 4$ is that corresponding to $\beta = 2$ in (\ref{1.1}) and
(\ref{1.2}). To obtain from the above results expressions corresponding
to the weight functions as written in (\ref{1.1}) and (\ref{1.2}) we must
make the replacements 
$$
S_4^{\rm hard}(X,Y) \mapsto 2  S_4^{\rm hard}(4X,4Y) \Big |_{a \to
2 a} , \quad S_4^{\rm soft}(X,Y) \mapsto {1 \over 2^{1/3}}
S_4^{\rm soft}(2^{2/3} X, 2^{2/3} Y).
$$

\section*{Acknowledgement}
The financial support of the ARC is acknowledged.

\section*{Appendix A}
\renewcommand{\theequation}{A.\arabic{equation}}
\setcounter{equation}{0}

In the case $\tau_{x} = \tau_{y} = 0$, 
the formula (\ref{3.8}) for $S_{2}$ at $\beta = 1$ 
reads 
\begin{equation}
\label{A.1}
S_2(x,y;0,0) = \sqrt{w(y)} \Big (
{\Phi_{N-2}(x;0) \over r_{N/2 - 1}(0)} - \sqrt{w(x)} {C_{N-1}(x) \over
h_{N-1}} \Big ) \Big (C_{N-1}(y) - (N-1) R_{N-2}(y) \Big ).
\end{equation}
Use of the explicit formulas (\ref{3.1}) and (\ref{3.2}) gives 
\begin{equation}
\label{A.2}
C_{N-1}(y) - (N-1) R_{N-2}(y)  = - (N-1)!
\Big ( L_{N-1}^a(y) - {d \over dy} L_{N-1}^a(y) \Big )
 =  (N-1)! {d \over dy} L_N^a(y),
\end{equation}
where the second equality  follows from an appropriate Laguerre polynomial 
identity \cite{Sz}. 

For the $x$-dependent factor in (\ref{A.1}), we require an explicit formula for 
$\Phi_{N-2}(x; 0)$. Now (\ref{2.1g}) substituted in (\ref{2.k}) gives 
\beqa
\nonumber
\Phi_k(x;0) = \int {\rm sgn}(x-y) \sqrt{w(y)} R_k(y).
\eeqa
Making use of the explicit formulas (\ref{3.2}) and (\ref{3.3}) then gives 
\beqa
\nonumber
{\Phi_{N-2}(x;0) \over r_{N/2 - 1}(0) } =
- {1 \over 4 \Gamma (a +N)} \int_0^\infty {\rm sgn}(x-y) y^{a/2} e^{-y/2}
{d \over dy} L_{N-1}^a (y) \, dy.
\eeqa
Integrating by parts, and making use of the Laguerre polynomial identity 
\beqa
\nonumber
(y - a) L_{N-1}^a(y) = y {d \over dy} L_{N-1}^a(y) - N
\Big ( L_N^a(y) - L_{N-1}^a(y) \Big ),
\eeqa
we see from this that 
\beqa
\label{A.3}\lefteqn{
{\Phi_{N-2}(x;0) \over r_{N/2 - 1}(0) }
=
\sqrt{w(x)} {C_{N-1}(x) \over h_{N-1}}
} \nonumber \\ && 
 +
{N \over 4 \Gamma (a+N)}  \int_0^\infty {\rm sgn}(x-y) y^{a/2-1} e^{-y/2}
\Big (L_N^a(y) -  L_{N-1}^a (y) \Big ) \, dy.
\eeqa
Substituting (\ref{A.2}) and (\ref{A.3}) in (\ref{A.1}) shows
\beqa
\label{A.4}\lefteqn{
S_2(x,y;0,0) = {N! \over 4 \Gamma(N+a)} \sqrt{w(y)}} \nonumber \\
&& \times
\Big ( {d \over dy} L_{N}^a(y) \Big )
 \int_0^\infty {\rm sgn}(x-y) y^{a/2-1} e^{-y/2}
\Big (L_N^a(y) -  L_{N-1}^a (y) \Big ) \, dy,
\eeqa
which is the formula recently given by Widom \cite{Wi98}. 

Consider next the expression (\ref{3.27}) for $S_{2}$ at $\beta = 4$ with 
$\tau_{x} = \tau_{y} = 0$. To simplify this expression, we note that the formula in 
(\ref{3.24}) for $R_{2j}(x)$ can be written in the simplified form 
\beqa
\label{A.5}
R_{2j}(x) = {1 \over a + 2j + 1}
\bigg ( C_{2j+1}(x) + {1 \over 2} (2j+1)! x^{-a/2} e^{x/2}
\int_0^x t^{a/2} e^{-t/2} \Big ( {d \over dt} L_{2j+2}(t) \Big ) \, dt
\bigg ).
\eeqa
This formula can be checked  by multiplying both sides by $x^{a/2}e^{-x/2}$ and 
differentiating. 
(We also verify that the $x \rightarrow \infty$ behaviour of both 
sides agrees.) 
Now, according to the definition (\ref{2.k}) of $\Phi_{k}$ and the 
second formula in (\ref{2.1g'}) 
\beqa
\nonumber
{\Phi_{N-2}(x;0) \over r_{N/2 - 1}(0) } =
- {2 \over r_{N/2 - 1}(0)} {d \over dx} \Big (
x^{a/2} e^{-x/2} R_{N-2}(x) \Big )
\eeqa
Substituting (\ref{A.5}), and making use of the differentiation formulas 
\beqa
\nonumber
y {d \over dy} L_n^a(y) = nL_n^a(y) - (n+a) L_{n-1}^a(y) =
(n+1) L_{n+1}^a(y) - (n+a+1-y)L_n^a(y)
\eeqa
and the fact that
\beqa
\nonumber
(a+N - 1) r_{N/2 + 1}(0) = h_{N-1} = \Gamma(N) \Gamma(N + a)
\eeqa
we readily find 
\beqa
\label{A.6}
{\Phi_{N-2}(x;0) \over r_{N/2 - 1}(0)} - \sqrt{w(x)}
{C_{N-1}(x) \over h_{N-1}} = - \sqrt{w(x)}
{N \over \Gamma(N + a)} {(L_N^a(x) - L_{N-1}^a(x)) \over x}
\eeqa
which thus simplifies the $x$-dependent factor in (\ref{3.27}). 
For the $y$-dependent 
factor, we see from (\ref{A.5}) that 
\beqa
\label{A.7}
\sqrt{w(y)}\Big ( C_{N-1}(y) - (a+N-1) R_{N-2}(y) \Big ) =
- {1 \over 2} (N-1)! \int_0^y t^{a/2} e^{-t/2} {d \over dt} L_N^a(t) \, dt.
\eeqa
Multiplying (\ref{A.6}) and (\ref{A.7}) shows
\beqa
\label{A.8}
S_2(x,y;0,0) = {N! \sqrt{w(x)} \over 2 \Gamma(N + a)}
{L_N^a(x) - L_{N-1}^a(x) \over x} \int_0^y t^{a/2} e^{-t/2}
{d \over dt} L_N^a(t) \, dt,
\eeqa
in agreement with the formula given recently by Widom \cite{Wi98}.

\end{document}